\documentclass[aip,jcp,reprint]{revtex4-1}
\usepackage{graphicx}
\usepackage{amsmath,amsxtra,amssymb,latexsym, amscd,color}
\usepackage[colorlinks=true, linkcolor=blue, urlcolor=blue, citecolor=blue]{hyperref}

\newcommand{\TXH}[1]{\textcolor{black} {#1}}

\begin{document}

\title{Protein escape at the ribosomal exit tunnel: Effects of native
interactions,  tunnel length and macromolecular crowding}

\author{Phuong Thuy Bui}
\affiliation{Duy Tan University, 254 Nguyen Van Linh, Thanh Khe, Da Nang, Viet Nam}

\author{Trinh Xuan Hoang}
\email{hoang@iop.vast.vn}
\affiliation{Institute of Physics, Vietnam Academy of Science and Technology,
10 Dao Tan, Ba Dinh, Hanoi, Vietnam}
\affiliation{Graduate University of Science and Technology, Vietnam Academy of
Science and Technology, 18 Hoang Quoc Viet, Cau Giay, Hanoi, Vietnam}

\date{\today}

\begin{abstract}
How fast a post-translational nascent protein escapes from the ribosomal
exit tunnel is relevant to its folding and protection against aggregation.
Here, by using Langevin molecular dynamics, we show that \TXH{non-local} native
	interactions help decreasing the escape time, and foldable proteins
	\TXH{generally escape much
	faster than same-length self-repulsive homopolymers at low
	temperatures. The escape process, however, is slowed
	down by the local interactions that stabilize the $\alpha$-helices.}	
The escape time is found to increase with both the tunnel length and the
concentration of macromolecular crowders outside the tunnel. 
We show that a simple diffusion model described by the Smoluchowski equation
with an effective linear potential can be used to map out the escape time
distribution for various tunnel lengths and various crowder concentrations. The
consistency between the simulation data and the diffusion model however is
found only for the tunnel length smaller than a cross-over length of 
90~{\AA} to 110~{\AA}, above which the escape time increases much faster with
the tunnel length. It is suggested that the length of ribosomal exit
tunnel has been selected by evolution to facilitate both the efficient folding
and efficient escape of single domain proteins. We show that macromolecular
crowders lead to an increase of the escape time, and attractive crowders
are unfavorable for the folding of nascent polypeptide.
\end{abstract}

\maketitle

\section{Introduction}

Partially folded protein conformations can be found at the ribosome during
protein translation and after translation before the full release of a nascent
protein from the ribosomal exit tunnel. The folding during translation, namely
cotranslational folding, occurs when a nascent polypeptide chain undergoes
elongation due to biosynthesis (for reviews, see e.g.
Refs.\cite{Fedorov1997,Cabrita2010,Cavagnero2011}), while the
post-translational folding is associated with a full-length protein. In both
cases, the behavior of the nascent polypeptide is strongly influenced by the
ribosome, especially by the ribosomal exit tunnel through which the nascent
chain traverses to the cytosol or to another cellular compartment.
The length of the ribosomal exit tunnel spans from 80~{\AA} to 100~{\AA},
depending on where the exit end is defined, whereas its width varies between
10~{\AA} and 20~{\AA} \cite{Voss2006}. Such a geometry allows for the formation
of an $\alpha$-helix or a $\beta$-hairpin inside the tunnel \cite{Deutsch2005},
and also promotes the $\alpha$-helix formation \cite{Thirumalai2005}, but would
hardly accommodate even a small tertiary structure \cite{Deutsch2009}.  The
tunnel was also suggested to have a recognitive
function leading to a translation arrest of certain amino acid sequences, such
as the SecM sequence \cite{Ito2002}. Cotranslational folding has been
characterized with vectorial folding \cite{Fedorov1997},
i.e. the folding events proceed from the N- to the C-terminus; the
non-equilibrium effect of a growing chain \cite{HoangPRL2005}; and the impact
of the varying codon-dependent translation rate
\cite{Zhang2009,Siller2010,OBrien2014,Nissley2014}. Furthermore, folding of 
nascent proteins is assisted by the action of ribosome-associated
molecular chaperones \cite{Frydman2001,Deuerling2005}. All these effects are
indicative of a highly conditional and coordinated folding of
nascent protein at the ribosome, which is clearly different from refolding
\cite{Anfinsen1972} of a denatured protein in aqueous solvent. There have been
experiments \cite{Clark2008,Clark2010,Bustamante2011} as well as simulations
\cite{Cieplak2015,Cieplak2015b,Thuy2016} showing that the folding efficiency of
proteins is improved under biosynthesis conditions. It was also suggested that
the impact of cotranslational folding is evolutionarily imprinted on the
protein native states, as seen with an increased helix propensity
\cite{HoangPRL2005} and a decreased compactness \cite{Alexandrov1993} of the
chain near the C-terminus in the statistical analyses of protein 
structures from the protein data bank (PDB). 

While cotranslational folding is progressively understood, little is known
about post-translational folding at the ribosome. The latter is considered
to take place after the protein C-terminus is released from the peptidyl
transferate center (PTC), where the peptide bonds are formed. Certainly,
protein must escape from the ribosomal tunnel to fully acquire the native
conformation. A too slow escape would decrease the productivity of the
ribosome, while a too fast escape would make the nascent protein vulnerable to
aggregation \cite{Dobson2003}, as the partially folded protein may still have a
large exposure of hydrophobic segments.
In a recent study \cite{Thuy2016}, by using MD simulations, we have
shown that post-translational folding at the exit tunnel is concomitant with
the escape process and that the tunnel induces a vectorial folding of the
full-length protein. Such a folding has a greatly reduced number of pathways
and leads to an improved folding efficiency. Interestingly, it has been also
shown \cite{Thuy2016} that the escape time distribution of protein can be
captured by a simple one-dimensional diffusion model of a particle in a
linear potential field with an exact solution of the Smoluchowski equation.

The purpose of the present study is to explore the protein escape at the 
ribosomal exit tunnel with a focus on several effects, namely the role of
native interactions, the impact of tunnel length and the influence of
macromolecular crowders \cite{Minton2001,Zhou2008}. We use the same approach as
given in our previous work \cite{Thuy2016}, that is to consider simple
coarse-grained models for the protein, the exit tunnel and the crowders, which
enable multiple simulations of protein growth and escape by using the Langevin
equation. The diffusion model for protein escape previously
introduced \cite{Thuy2016} is improved in this study by considering an
absorbing boundary
condition. We find that escape time reflects well the changes in the system
properties, such as the native contact map, the tunnel length and the
crowder concentration with a remarkable consistency between simulations
and the theoretical diffusion model. Interestingly, the dependence of 
the escape time on the tunnel length suggests an explanation for the observed
length of the exit tunnel in real ribosomes. Our results obtained with
attractive crowders provide an insight into the effect of ribosome-associated
chaperones on the escape and folding of nascent proteins at the ribosome.

\section{Methods}

\subsection{Models of nascent protein, ribosomal exit tunnel and macromolecular
crowders}

As a nascent protein we will focus on the B1 domain of protein G of length
$N=56$ amino acids with the PDB code of 1pga, denoted as GB1. The protein is 
considered in a Go-like model
\cite{Go1983,HoangJCP2000,HoangJCP2000b,Clementi2000}, in which each amino acid
is considered a single bead centered at the position of the
C$_\alpha$ atom. We adopt the same Go-like model as given in our previous
work \cite{Thuy2016} except that with a 10-12 Lennard-Jones (LJ) potential
for native contact interactions. In addition, we consider three types of native
contact maps, denoted as C1, C2 and C3, for the model.
The C1 map is defined by a cut-off distance of 7.5~{\AA}
between the C$_\alpha$ atoms in the native conformation.  The C2 and C3 maps
are obtained based on an all-atom consideration \cite{Cieplak2002} of the
protein PDB structure: contact between two amino acids is identified if there
are least two non-hydrogen atoms belonging to the two amino acids, found at a
distance shorter than $\lambda$ times the sum of their atomic van der Waals
radii. C2 map has $\lambda=1.27$ whereas C3 has $\lambda=1.5$.  The choice of
$\lambda=1.27$ is such that the C2 map has the same number of native contacts
and the C1 for the GB1 protein.  The interaction
between a pair of amino acids forming a native contact takes the form of a
10-12 LJ potential \cite{Clementi2000}
\begin{equation}
V(r_{ij}) = \epsilon\left[5(r_{ij}^*/r_{ij})^{12} - 
6(r_{ij}^*/r_{ij})^{10}\right] \ ,
\label{eq:v1012}
\end{equation}
where $\epsilon$ is an energy unit in the system corresponding to the strength
of the LJ potential, $r_{ij}$ is the distance between 
residues $i$ and $j$, and $r_{ij}^*$ is the corresponding distance in the
native state. The use of 10-12 LJ potential makes the folding transition
more cooperative \cite{Kaya2000} than the 6--12 LJ potential (used in 
previous work \cite{Thuy2016}) as indicated by the height of the specific heat
peak (see Fig. S1 of the supplementary material).
\TXH{Additionally, we will consider also a number of small single-domain
proteins to study the effects of native interactions and the tunnel length
on the escape process.}

\begin{figure}
\includegraphics[width=3.4in]{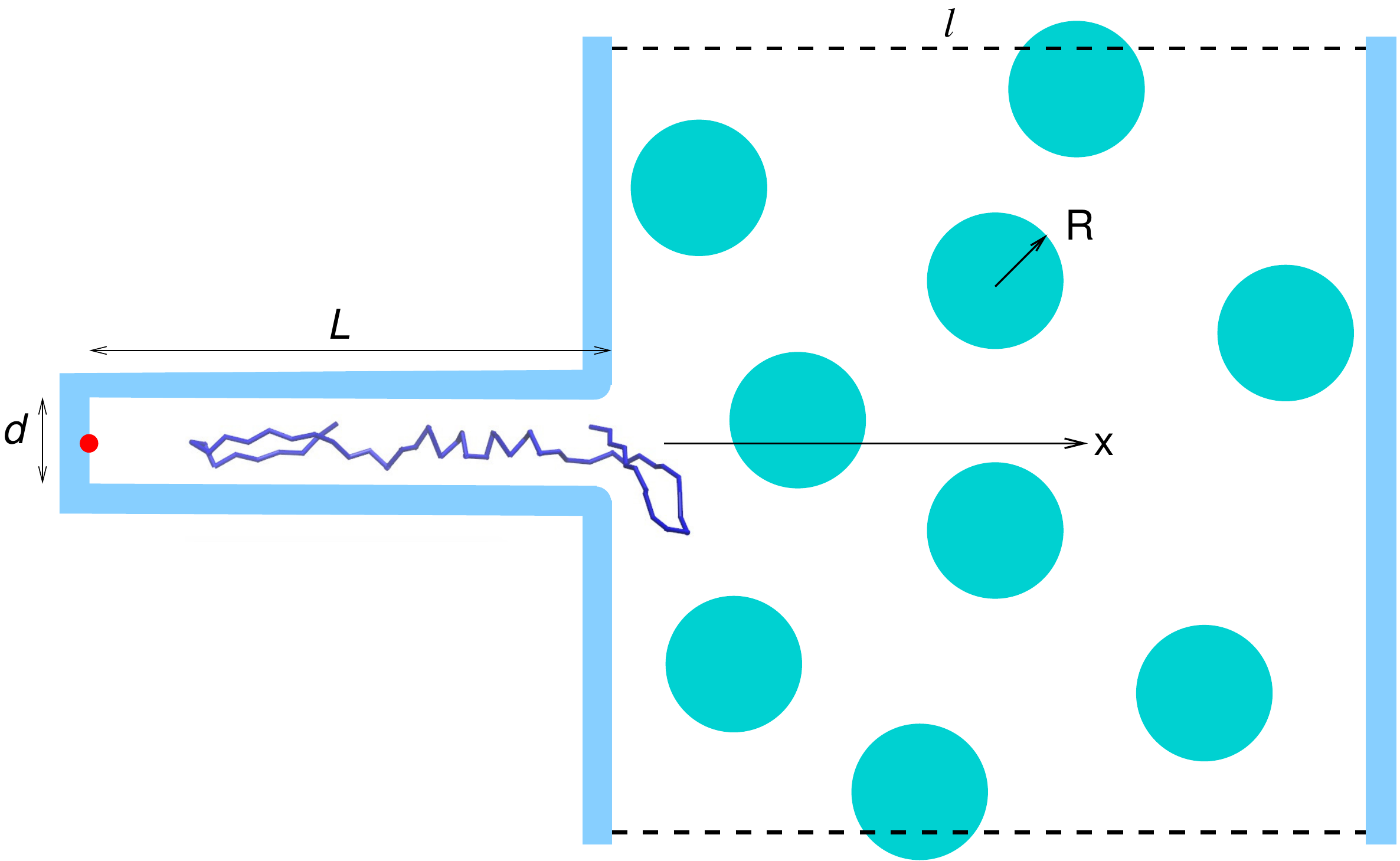}
\caption{Sketch of the models of a ribosomal exit tunnel with a partially
folded nascent protein inside and macromolecular crowders outside the tunnel.
The peptidyl transferate center (PTC), where the protein is grown, is shown as
a red circle.} 
\label{fig1}
\end{figure}

The ribosomal tunnel is modeled as a hollow cylinder of repulsive walls with
diameter $d=15$~{\AA} and length $L$ (Fig.  \ref{fig1}). It has been shown
\cite{Thuy2016} that this diameter allows for the formation of an
$\alpha$-helix and a $\beta$-hairpin inside the tunnel but not tertiary
structures. In the present study, $L$ is allowed to changed between 0 and
140~{\AA}. The cylinder
has one of its circular bases open and attached to a repulsive flat wall
mimicking the ribosome's outer surface.  Macromolecular crowders are modeled as
\TXH{soft spheres} of radius \TXH{$R=10$ {\AA}} \TXH{($R$ is chosen
approximately equal to the radius of gyration of GB1, $R_g=10.2$ {\AA})}.
Assume that the $x$ axis is the tunnel axis, the crowders are confined between
the ribosome's wall and another wall parallel to it at a distance $l=100$~{\AA}
along the $x$ direction. Periodic boundary conditions are applied for the $y$
and $z$ directions with a box size equal to $l$. The crowders' volume fraction
is given by $\phi=M (4\pi/3) R^3/l^3$, with $M$ the number of crowders.

\begin{figure}
\includegraphics[width=3.2in]{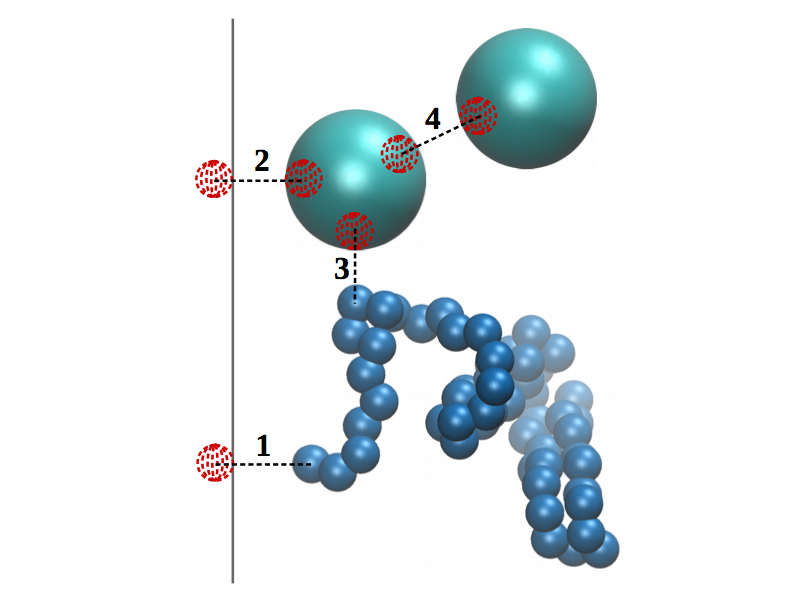}
	\caption{\TXH{Virtual residues (red dotted spheres) in the interactions
	between an amino acid and a wall (1), between a crowder and a wall (2),
	between an amino acid and a crowder (3), and between two crowders
	(4). The virtual residues are of the same size as amino acid (blue) 
	and can be at any position embedded under the surface of a wall
	(solid line) or a crowder (cyan). The interaction potentials involving
	a wall or a crowder are defined based on the distance to the nearest
	virtual residue (as in 1 and 3) or between the two nearest virtual
	residues (as in 2 and 4) belonging to these objects.}} 
	\label{virtual}
\end{figure}

The interactions between an amino acid and a wall, between a crowder and a
wall, and between two crowders are all repulsive and given in the form of a
shifted and truncated LJ potential
\begin{equation}
V_\mathrm{rep}(r) = \left\{ \begin{array}{ll}
 4 \epsilon \left[(\sigma/r)^{12} - (\sigma/r)^6 \right]
 + \epsilon &, \quad r \leq 2^{1/6} \sigma \\
0 &, \quad r > 2^{1/6} \sigma
\end{array} \right. \ ,
\label{eq:vrep}
\end{equation}
where $\sigma=5$~{\AA} is a characteristic length equal to the typical diameter
of an amino acid; $r$ is the distance between an amino acid and the nearest
virtual residue \cite{Thuy2016} of diameter $\sigma$, embedded under the
surface a wall or a crowder, or between two such virtual residues \TXH{(see
Fig. \ref{virtual} for the definition of virtual residue)}.

Additionally, we will consider also a case in which the crowders are weakly
attractive to amino acids with the interaction given by the 6--12 LJ potential
\begin{equation}
V_\mathrm{att}(r) = 4 \epsilon_1 \left[(\sigma/r)^{12} - (\sigma/r)^6 \right]  
\ ,
\end{equation}
with energy parameter $\epsilon_1 < \epsilon$.

\begin{figure}
\includegraphics[width=3.4in]{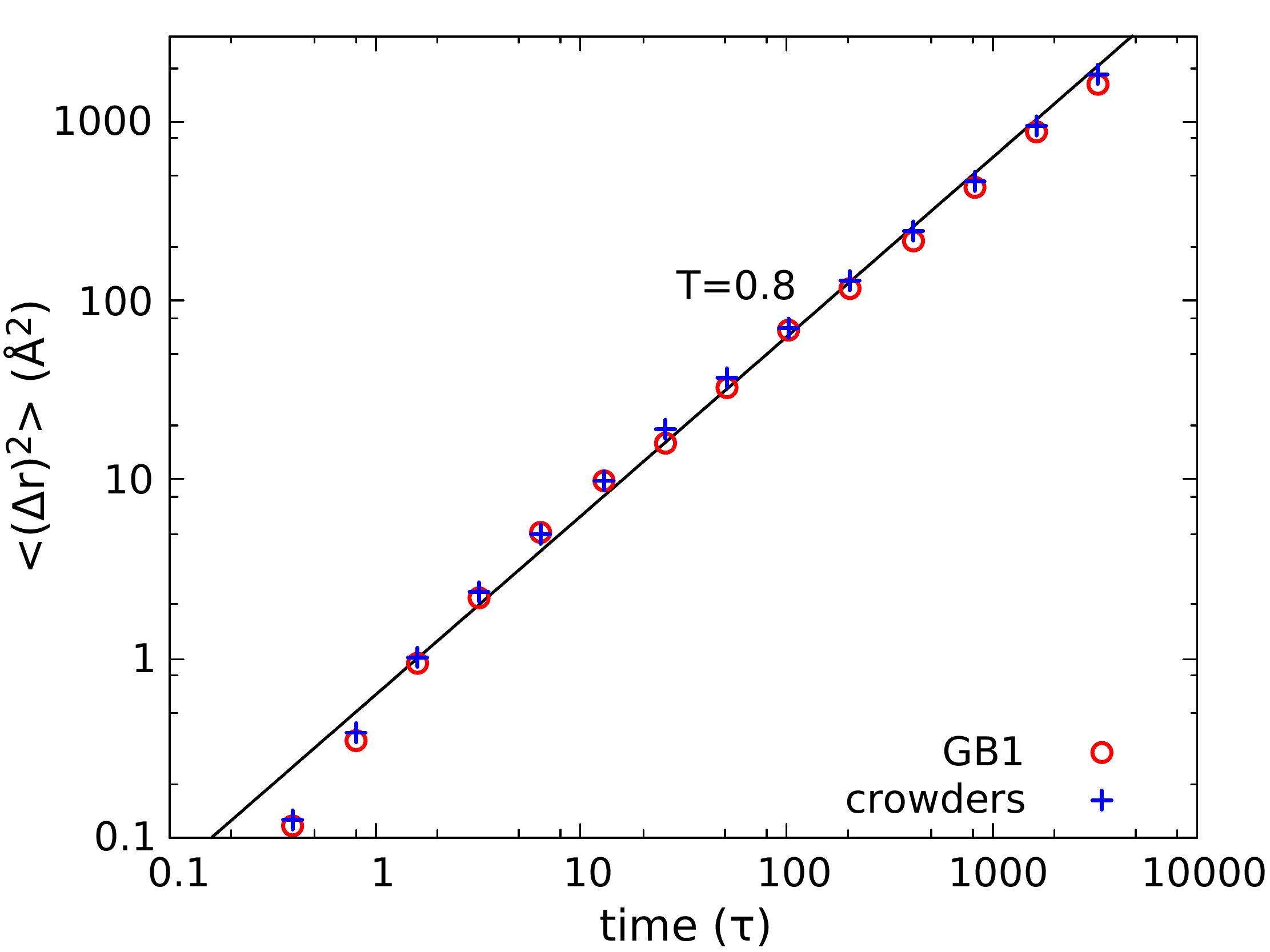}
\caption{ 
Time dependence of the mean square displacement of the protein GB1 (circles)
and crowders (crosses) in simulations without the tunnel. The simulations
are carried out for the protein and 47 crowders at the volume fraction
of $\phi=0.2$. The simulation temperature is $T=0.8\,\epsilon/k_B$, at which
the protein stays in its native state. The protein displacement is calculated
using its center of mass. The averages are taken over 100 independent
trajectories. The solid line has a slope equal to 1. 
	} 
\label{fig2}
\end{figure}

The motions of the protein and the crowders are simulated by using the Langevin
equations \cite{Thuy2016,HoangJCP2000}. One assumes that all amino acids have
the same mass, $m$, while the crowders has a molecular mass $m_c$ equal to the
mass of the protein, i.e. $m_c = N m$.  Similarly, the friction coefficient of
amino acid is $\zeta_a$, whereas that of crowder is $\zeta_c = N \zeta_a$. The
Langevin equations are integrated by using a Verlet algorithm introduced in
Ref. \cite{Thuy2016} with time step $\Delta t = 0.002 \tau$, where $\tau =
\sqrt{m \sigma^2/\epsilon}$ is the time unit in the system. In the simulations
we use $\zeta_a = 5 \, m \tau^{-1}$, for which the dynamics of the system are
in the overdamped limit \cite{Thuy2016,Klimov1997}.  Fig. \ref{fig2} shows that
the employed dynamics lead to the same diffusion characteristics for a folded
protein and for crowders in the solution. These characteristics are also close
to that of a Brownian motion for times larger than $\tau$. 

Following previous work \cite{Thuy2016}, the nascent protein is grown inside
the tunnel at the PTC from the $N$ terminus to the $C$ terminus with the growth
time per amino acid $t_g = 100 \tau$. This growth speed is sufficiently slow
to produce fully translated conformations of similar structural characteristics
to those obtained by a much slower growth speed \cite{Thuy2016}. The escape
time is measured from the moment the protein has grown to its full length until
all of its amino acids are escaped from the tunnel.

\subsection{Diffusion model of protein escape}

\TXH{Our previous study \cite{Thuy2016} has shown that the protein escape at
the exit tunnel in absence of crowders is a downhill process corresponding
to a free energy which monotonically decreases along a reaction coordinate
associated with the escape degree, such as the number of residues outside the
tunnel or the position of the C-terminus. Such a process is consistent with the
diffusion of a particle in an one-dimensional external potential field $U(x)$
where $U$ is a decreasing function of $x$.} This diffusion process is described
by the Smoluchowski equation
\cite{vankampen}
\begin{equation}
\frac{\partial}{\partial t}\, p(x,t|x_0,t_0) =  
\frac{\partial}{\partial x} D
 \left(\beta \frac{\partial U(x)}{\partial x} + \frac{\partial}{\partial x}
\right)\, p(x,t|x_0,t_0) ,
\label{eq:smolu}
\end{equation}
where $p(x,t|x_0,t_0)$ is a conditional probability density of finding the
particle at position $x$ and at time $t$, given that it was found previously at
position $x_0$ at time $t_0$; $D$ is diffusion constant, assumed to be position
independent; and $\beta=(k_B T)^{-1}$ is the inverse temperature with $k_B$ the
Boltzmann constant. Assume that the external potential field has a linear form,
$U(x) = - k x$, with $k$ a constant. 
\TXH{For a nascent protein at the tunnel, the constant $k$ presents an average
slope of the dependence of the free energy of the protein on the escape
coordinate.}
In such a case, a solution of Eq.
(\ref{eq:smolu}) for an unconstrained particle is given by
\begin{equation}
p(x,t) \equiv p(x,t|0,0)=\frac{1}{\sqrt{4\pi D t}} \exp\left[
-\frac{(x - D\beta k t)^2}{4Dt}
\right] .
\label{eq:pxt}
\end{equation}
given that the initial condition is $p(x,0)=\delta(x)$.
This solution gives the mean displacement of the particle
\begin{equation}
\langle x \rangle = (D\beta k) t \ ,
\end{equation}
with a diffusion speed equal to $D\beta k$.
\TXH{For a Brownian particle, $D$ depends on the temperature $T$ and on the
friction coefficient $\zeta$ according to the Einstein's relation
\begin{equation}
	D = \frac{k_B T}{\zeta} \ .
	\label{eq:einstein}
\end{equation}}

The escape time of nascent protein at the tunnel corresponds to the first
passage time (FPT) of a diffused particle subject to the initial condition at
$x=0$ and an absorbing boundary condition at $x=L$. The latter condition is
given as
\begin{equation}
p(L,t) = 0\ .
\end{equation}
The FPT distribution for this absorbing boundary condition can be found
in Ref. \cite{Cox} and is given by
\begin{equation}
g (t)=\frac{L}{\sqrt{4\pi D t^3}} \exp\left[
-\frac{(L - D\beta k t)^2}{4Dt}
\right] .
\label{eq:pt}
\end{equation}
Using the distribution in Eq. (\ref{eq:pt}) one obtains the mean escape
time 
\begin{equation}
\mu_t \equiv \langle t \rangle = \int_0^\infty t \, g(t)\, dt 
= \frac{L}{D \beta k} \ ,
\label{eq:mut}
\end{equation}
and the standard deviation 
\begin{equation}
\sigma_t \equiv (\langle t^2 \rangle - \langle t \rangle^2)^{\frac{1}{2}}
= \frac{\sqrt{2 \beta k L}}{D (\beta k)^2} \ .
\label{eq:sigt}
\end{equation}
It follows that the ratio $\sigma_t/\mu_t$ is independent of $D$.  Note that
both $\mu_t$ and $\sigma_t$ diverges when $k=0$, for which $g(t)$ becomes the
heavy-tailed L\'evy distribution. 
It can be expected that $D$ and $\beta k$ may depend on $L$ and on the
crowders' volume fraction $\phi$. These dependences will be investigated in the
present study. 

\section{Results and Discussion}

\subsection{Effect of native interactions}

Our previous study \cite{Thuy2016} has shown that the folding of nascent
proteins speeds up their escape process at the ribosomal tunnel. Here, we
investigate how this enhancement is sensitive to the details of native
interactions and how the escape of a protein is different from that of a
homopolymer. For this investigation, we fix the length of the tunnel to be
$L=80$~{\AA} and consider the protein without crowders.

We consider the protein GB1 with three different native contact maps, C1, C2
and C3, as described in the Methods section.  Both the C1 and C2 maps have 102
native contacts, but of which only 72 contacts are common. The C2 map has more
long-range contacts than the C1 one. The \TXH{relative contact order (CO)}
\cite{Plaxco1998} of the C2 map ($\approx 0.3444$), is higher than that of the
C1 map ($\approx 0.3283$). The C3 map has 120 contacts (\TXH{CO} $\approx
0.3509$) and includes all the contacts in the C2 map. The folding temperature
$T_f$ of a free protein without the tunnel is defined as the temperature of the
maximum of the specific heat peak (Fig. S1 of the supplementary material) and
equal to 0.866, 0.888 and 1.004 $\epsilon/k_B$ for the models with C1, C2 and
C3 contact maps, respectively.  We consider also two
homopolymers of the same length as the GB1 protein ($N=56$). The first one is a
self-repulsive homopolymer with a repulsive potential of
$\epsilon\,(\sigma/r)^{12}$ for the interaction between any pair of
non-consecutive beads. The second homopolymer is a self-attractive one with the
12-10 LJ potential, given by Eq.  (\ref{eq:v1012}), for the attraction between
the beads. Note that the self-repulsive homopolymer can be considered as
representing an intrinsically disordered protein, with regards to an important
class of proteins that do not fold in vivo \cite{Dyson2015}.

\begin{figure}
\includegraphics[width=3.4in]{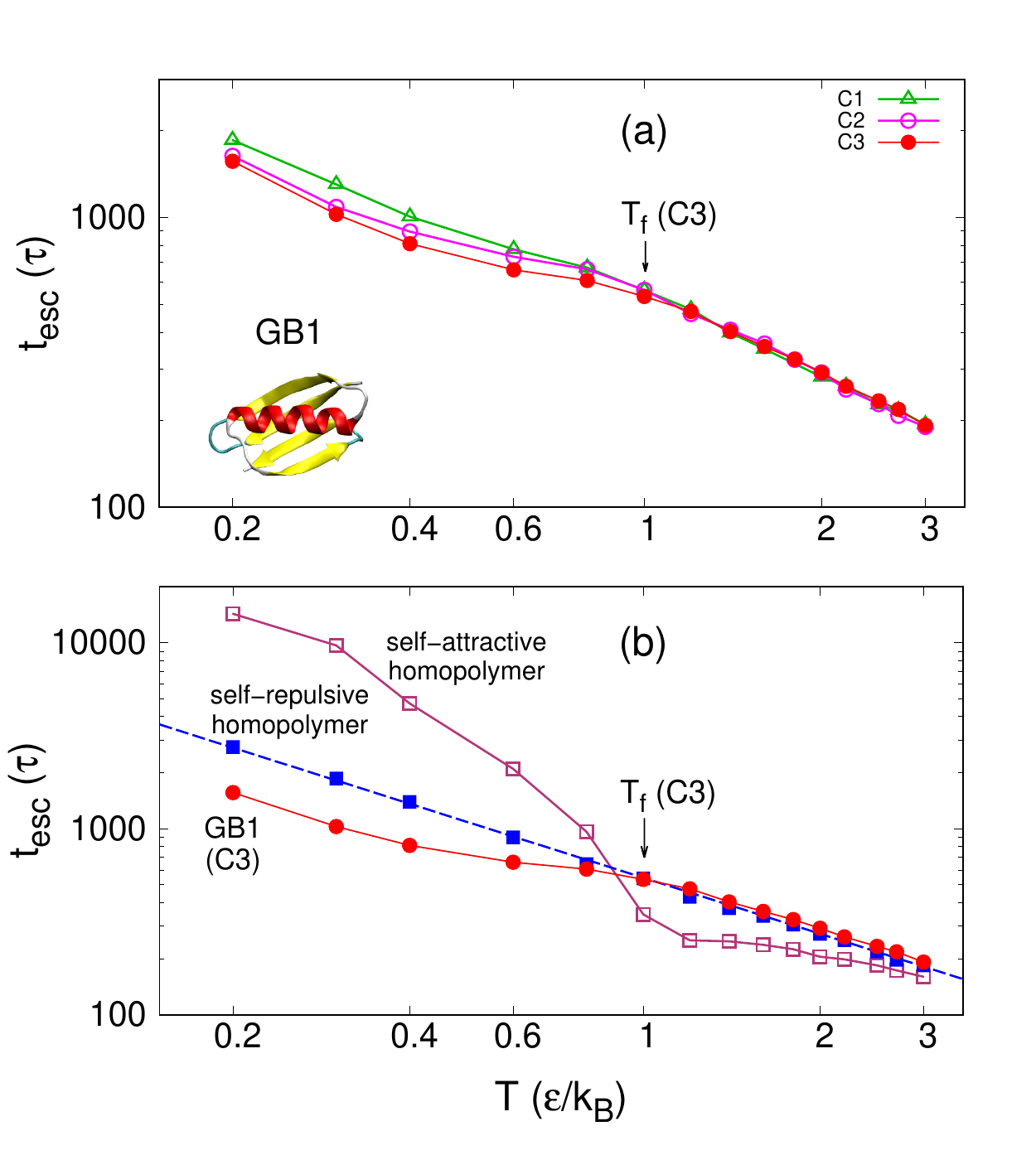}
\caption{Dependence of the median escape time, $t_\mathrm{esc}$, on
temperature, $T$, at the tunnel of length $L=80$~{\AA}: (a) For the GB1
protein in Go-like models with the C1 (triangles), C2 (open circles) and
C3 (filled circles) native contact maps; (b) For GB1 with the C3 map (filled
circles), the self-repulsive homopolymer (filled squares) and the
self-attractive homopolymer (open squares). The escape times for the
self-repulsive homopolymer are fitted with a $T^{-1}$ dependence (dashed line).
Arrow indicates the folding temperature $T_f=1.004\,\epsilon/k_B$ for the
protein with the C3 map.
	\TXH{The native state of GB1 is shown as inset in (a).}
} 
\label{fig3}
\end{figure}

Fig. \ref{fig3}a shows the dependence of the median escape time,
$t_\mathrm{esc}$, on temperature for the GB1 protein with the three native
contact maps. It is shown that for temperatures roughly larger than $T_f$, all
three contact maps lead to almost the same escape times. For $T<T_f$,
differences in the escape times are found among the models, even though the
dependences of $t_\mathrm{esc}$ on $T$ are of similar shape for all the three
models. The model with the C3 map has the smallest escape times, indicating
that the larger number of native contacts the faster is the escape of the
protein. On the other hand, the model with C2 map has smaller $t_\mathrm{esc}$
than the model with the C1 map, despite that they have the same number of
native contacts. This result indicates that the escape time also depends on the
details of the native contact map, and a protein with more long-range contacts
would have a faster escape from the tunnel. 

Fig. \ref{fig3}b compares the escape times of the GB1 protein with the C3
native contact map with the two homopolymers. It shows that for $T>T_f$, the
protein has the escape time \TXH{slightly larger but} close to that of the
self-repulsive homopolymer, as expected for an unfolded chain. For $T<T_f$, the
protein escapes faster than the homopolymer with self-repulsion, reconfirming
the favorable effect of
folding on the escape process. The self-attractive homopolymer shows a
very different behavior of the escape time than the self-repulsive one. In
particular, for temperatures lower than an intermediate temperature of about
$0.9\,\epsilon/k_B$, the self-attractive homopolymer has a much larger escape
time than the self-repulsive one; while an opposite trend is seen for
$T>0.9\,\epsilon/k_B$, for which the self-attractive polymer escapes faster
than the other one. We find that for $T < 0.9\,\epsilon/k_B$, 
the full-length homopolymer starts the escape process with a collapsed
conformation completely fitted inside the tunnel. From this conformation, the
polymer diffuses very slowly in the tunnel until a part of it emerges from
the tunnel. For $T > 0.9\,\epsilon/k_B$, the polymer begins to escape with a
conformation having a small part found outside the tunnel. We have checked that
the self-attractive homopolymer has a collapse transition temperature $\approx
2.2\,\epsilon/k_B$, thus below this temperature but above 0.9 $\epsilon/k_B$,
its escape process is accelerated by the collapse of the chain.  Above the
collapse transition temperature, the escape time of the self-attractive
homopolymer is smaller but approaching that of the self-repulsive one as
temperature increases.
\TXH{Note that below the collapse transition temperature, the size of the
collapsed polymer still depends on temperature. Thus, the temperature of 0.9
$\epsilon/k_B$, at which a rapid change in the escape time is seen, should
be understood as specific to the polymer length and the tunnel length
considered. At this temperature, the typical size of the self-attractive
homopolymer along the tunnel axis approximately matches that of the tunnel.}

The result of the self-attractive homopolymer shows that the collapse of
the chain accelerates the escape process only when the chain has a part
found outside the tunnel. It indicates the relative size of the polymer
to the tunnel length and also temperature are relevant to the escape behavior
of the polymer. We find that protein behaves similarly on increasing the
tunnel length, as will be shown in the next subsection. 

Only for the self-repulsive homopolymer, the escape time is proportional to
$T^{-1}$ for the whole range of temperature. As $\beta k$ is approximately
constant on changing temperature (see Fig. \ref{fig8} of this study and also
Ref. \cite{Thuy2016}), it follows from Eq. (\ref{eq:mut}) that the diffusion
coefficient $D$ of the self-repulsive homopolymer is proportional to $T$,
consistent with the Einstein's relation for Brownian particle (Eq.
(\ref{eq:einstein})). Deviation from this Brownian behavior on changing
temperature thus is observed for the protein and the self-attractive
homopolymer due to the fact that they adopt different compact conformations
during the escape process at temperatures below their folding or collapse
transition temperatures.

\begin{figure}
\includegraphics[width=3.4in]{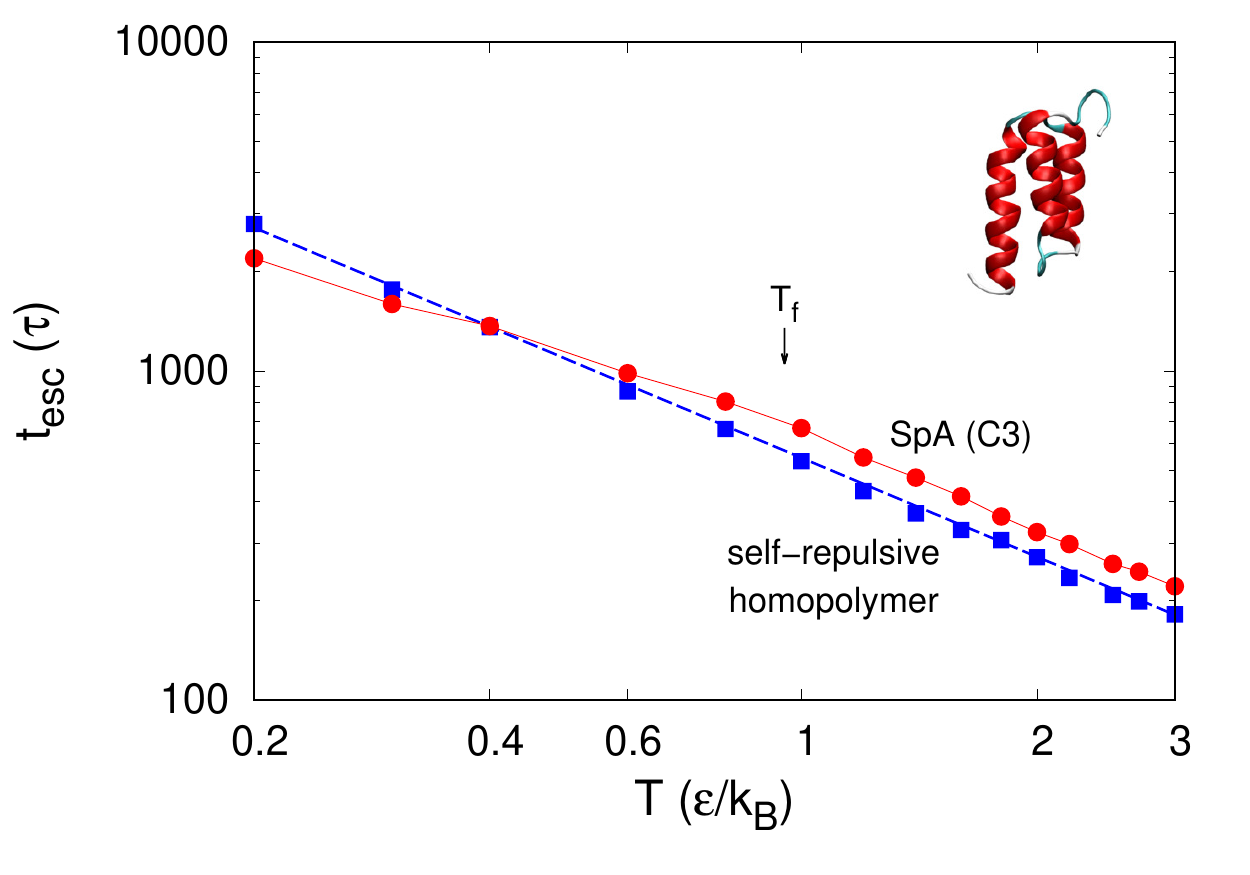}
\caption{
	\TXH{Dependence of the median escape time, $t_\mathrm{esc}$, on
	temperature, $T$, for the SpA protein (circles) and the self-repulsive
	homopolymer (squares). The homopolymer has the same length ($N=58$) as SpA.
	The SpA is considered in the Go-like model with the C3 native contact
	map, and its native conformation is shown as inset. The folding temperature
of SpA, $T_f=0.952 \,\epsilon/k_B$, is indicated by an arrow.}
}
	\label{figspa}
\end{figure}

\begin{figure}
\includegraphics[width=3.4in]{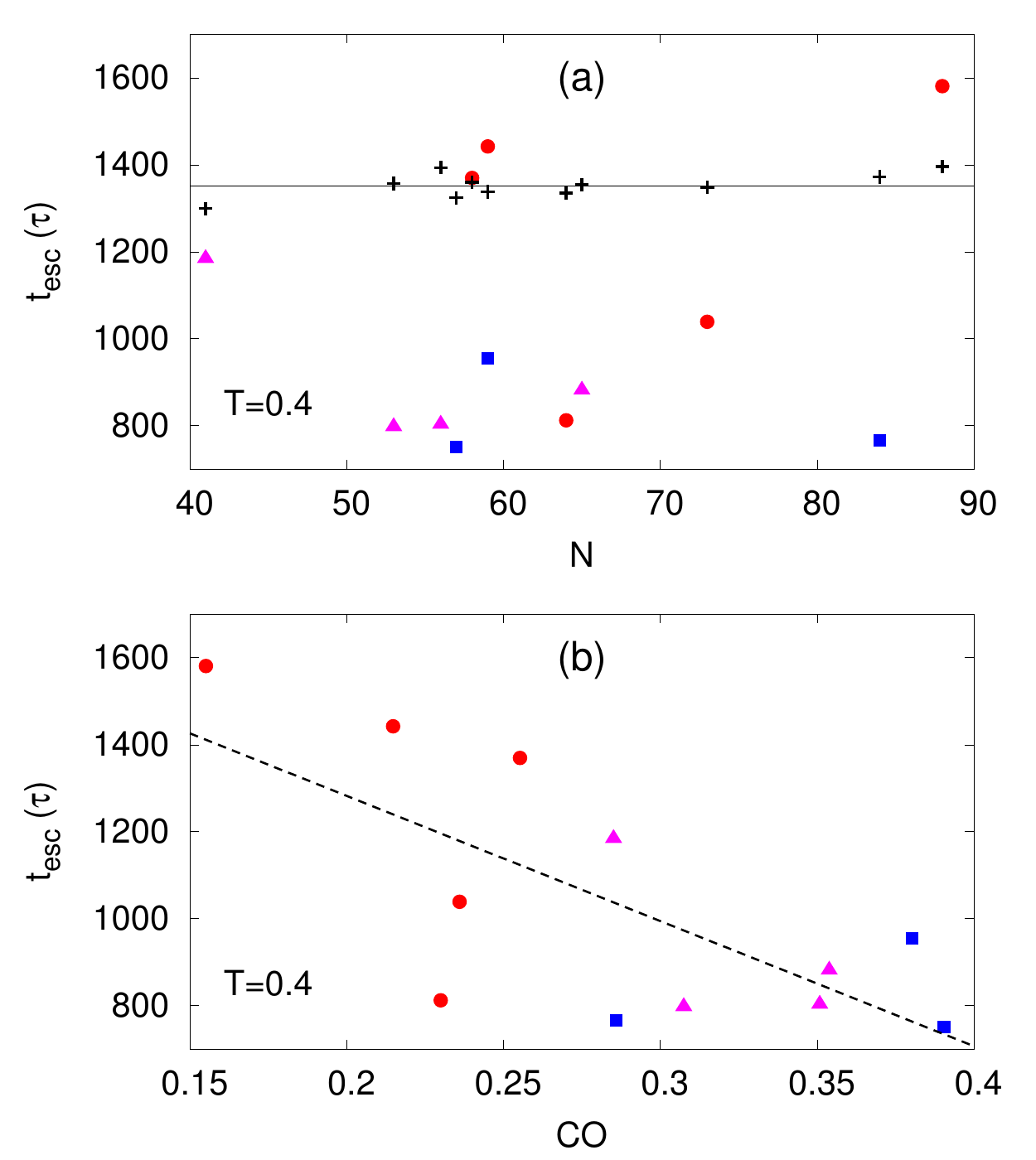}
\caption{\TXH{(a) Dependence of the median escape time, $t_\mathrm{esc}$,
on the chain length, $N$, of proteins (filled symbols) and the self-repulsive
homopolymers (crosses).  The data shown are obtained at $T=0.4\,\epsilon/k_B$
for 12 small single-domain proteins with PDB codes 1iur, 2jwd, 2rjy, 1wxl,
2spz, 1wt7, 2erw, 1pga, 2ci2, 1f53, 2k3b, and 1shg, classified as all-$\alpha$
(circles), $\alpha$/$\beta$ (triangles) and all-$\beta$ (squares), in the
Go-like model with the C3 native contact map, and for corresponding
homopolymers of the same lengths as the proteins. The average escape time of
the homopolymers is indicated by horizontal line.
(b) Dependence of $t_\mathrm{esc}$ on the relative contact order (CO) for
the proteins (filled symbols) with an average trend shown as dashed line.  } }
\label{figcorder}
\end{figure}

\TXH{We have calculated the escape time for a number of small single-domain
proteins other than GB1 with different native state topologies. Fig.
\ref{figspa} shows the dependence of the median escape time on temperature for
the Z domain of Staphylococcal protein (SpA) of length $N=58$. The native state
of SpA is a three-helix bundle (Fig. \ref{figspa}, inset). It is shown that at
high temperatures, the escape time of SpA is higher than the escape time of a
same-length self-repulsive homopolymer. However, for $T < T_f$, the relative
difference between the two escape times decreases with temperature, and for
$T<0.4\,\epsilon/k_B$, the protein escape faster than the homopolymer. Thus,
the folding of SpA enhances its escape process. We have found that
another helix bundle with the PDB code 2rjy also escapes faster then the 
homopolymer at temperatures lower than $T_f$ (see Fig. S2 of supplementary
material).  On the other hand, a single $\alpha$-helix escapes more
slowly than the self-repulsive homopolymer at all temperatures (see Fig. S3 of
supplementary material). These results suggest that local
interactions stabilizing the $\alpha$-helix slow down the escape process and the
latter is accelerated only by the non-local interactions.}

\TXH{Note that the Go-like model includes local potentials on the bond angles
and dihedral
angles favoring native conformation while the homopolymer model does not. Even
at temperatures higher than $T_f$, these interactions still have some effect on
the local conformations. For $\alpha$-helical proteins, they make the
escaping protein conformations less extended and more rigid than those of the
self-repulsive homopolymer. This effect explains why $\alpha$-helical proteins
escape more slowly than the self-repulsive homopolymer for 
$T>T_f$, at least for $T$ up to 3$\,\epsilon/k_B$ as shown in Fig.
\ref{figspa}.  It can be expected that for much higher temperatures, at which
the local potentials become unimportant, the escape time of protein approaches
that of the homopolymer.}

\TXH{The different effects of local and non-local interactions on the escape
time of proteins can also be seen in Fig. \ref{figcorder}. Fig.
\ref{figcorder}b shows that at a temperature favorable for folding,
$T=0.4\,\epsilon/k_B$, the escape time of protein to a considerable degree is
correlated with the relative contact order.
Fig. \ref{figcorder}a shows that the escape time of protein is uncorrelated
with the chain length, whereas that of the self-repulsive homopolymer is almost
independent on the chain length. Fig. \ref{figcorder}a also shows that the
$\alpha$/$\beta$ and all-$\beta$ proteins have smaller
escape time than the same-length homopolymers, whereas the all-$\alpha$ proteins
may have smaller or larger escape time than the homopolymers. 
We have checked that among the all-$\alpha$ proteins considered, the larger
the number of non-local contacts, the faster the protein escapes.
}

\subsection{Effect of tunnel length}

We study now the dependence of the escape time on the length of the
ribosomal tunnel, which is considered as an adjustable parameter in our model.
For this investigation, we have carried out simulations for the GB1 protein
with the tunnel length $L$ varied between 10 and 130~{\AA}, and analyzed the
statistics of the escape times using the insights from the diffusion model.
From here on, for simplicity, we consider only the Go-like model with the C3
native contact map for protein GB1.

\begin{figure}
\includegraphics[width=3.4in]{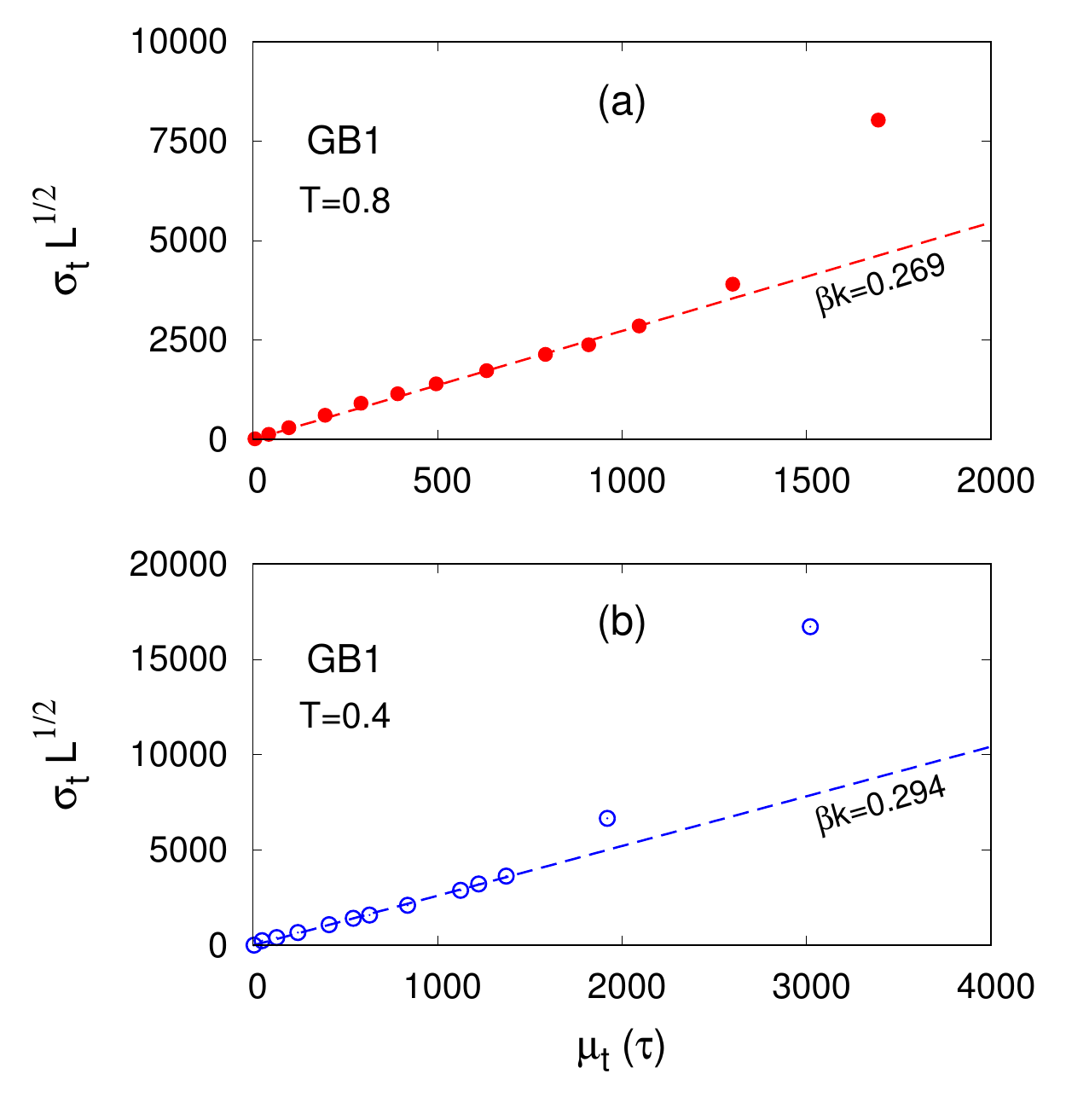}
\caption{Dependence of the standard deviation of the escape time multiplied by
the square root of the tunnel length, $\sigma_t L^{1/2}$, on the mean 
escape time, $\mu_t$, for protein GB1 with the C3 contact map at two
temperatures, $T=0.8\epsilon/k_B$ (a) and $T=0.4\epsilon/k_B$ (b). The data
points shown are obtained for various tunnel length $L$ between 10 and
130~{\AA}. The points associated with $L\leq 110$~{\AA} are fitted to a linear
function corresponding to the diffusion model with $\beta k=0.269$~{\AA}$^{-1}$
(a) and $\beta k = 0.294$~{\AA}$^{-1}$ (b).}
\label{fig4}
\end{figure}

The diffusion model predicts that the ratio between the standard deviation
of the escape time, $\sigma_t$, and the mean escape time, $\mu_t$, is
given by
\begin{equation}
\frac{\sigma_t}{\mu_t} = \sqrt{\frac{2}{L\beta k}} \ ,
\label{eq:ratio}
\end{equation}
and thus depends only on $L$ and $\beta k$. Fig. \ref{fig4} shows that the
dependence of $\sigma_t L^{1/2}$ on $\mu_t$ for GB1 obtained by the simulations
at two different temperature below $T_f$ is almost linear for $L \leq
110$~{\AA}. This linear dependence indicates that $\beta k$ is constant on
changing $L$, for $L \leq 110$~{\AA}. The fits of the simulation data to Eq.
(\ref{eq:ratio}) show that $\beta k = 0.269$~{\AA}$^{-1}$ for
$T=0.8\epsilon/k_B$ and $\beta k = 0.294$~{\AA}$^{-1}$ for $T=0.4
\epsilon/k_B$. Thus, the values of $\beta k$ are not the same but quite close
for the two temperatures considered. Fig. \ref{fig4} shows that for $L >
110$~{\AA}, the dependence of $\sigma_t L^{1/2}$ on $\mu_t$ strongly
deviates from the linear dependence obtained for smaller $L$, indicating
that $\beta k$ quickly decreases on increasing $L$. Thus, the diffusion
properties of the protein changes qualitatively at the tunnel length of $L
\approx 110$~{\AA}. We call the latter the cross-over length for
the diffusion of protein at the tunnel.

\begin{figure}
\includegraphics[width=3.4in]{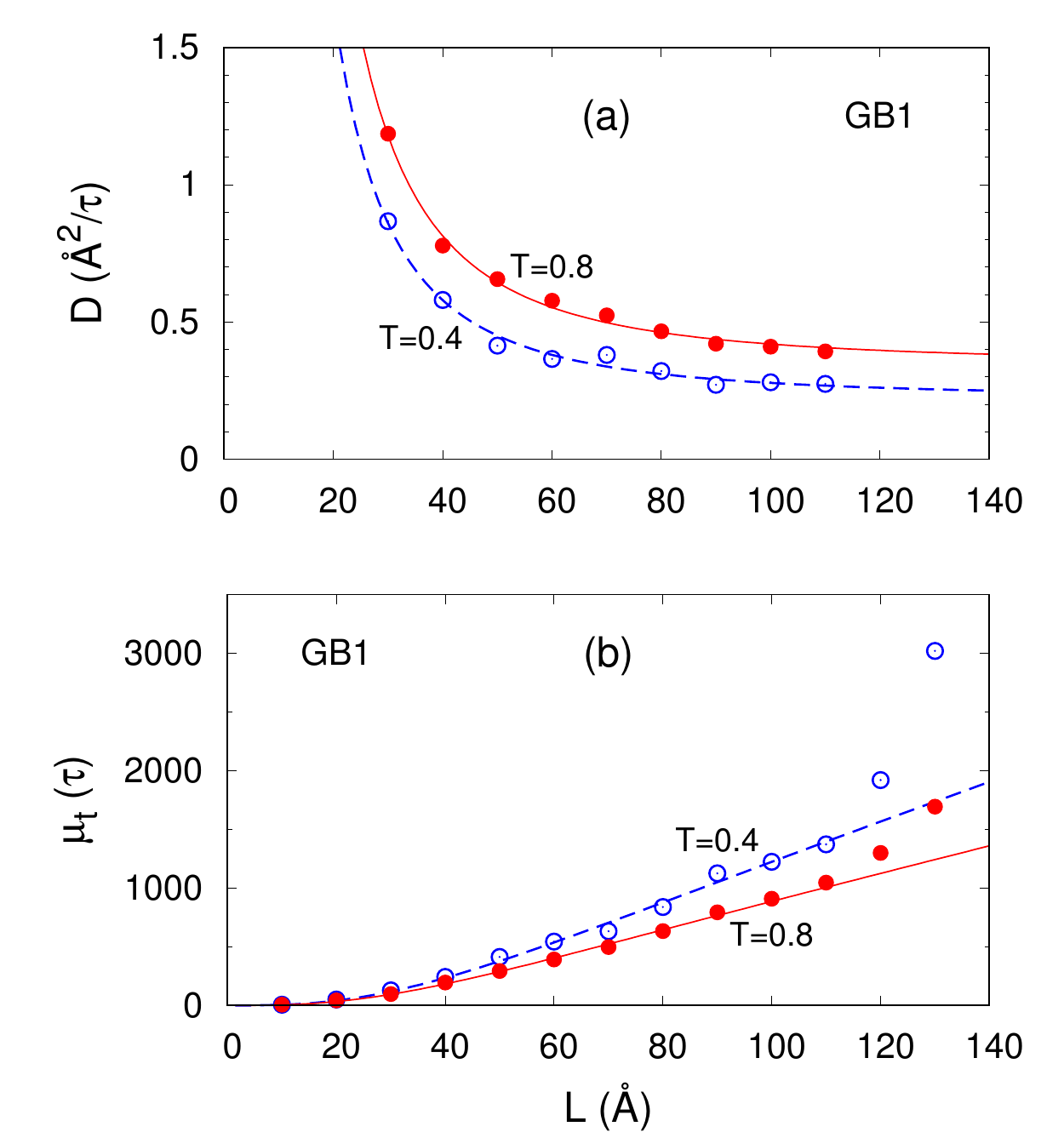}
\caption{
Dependence of the diffusion constant, $D$, (a) and the mean escape
time, $\mu_t$, (b) on the tunnel length, $L$, for the GB1 protein with
C3 contact map at $T=0.8\,\epsilon/k_B$ (filled circles) and $T=0.4
\,\epsilon/k_B$ (open circles). The values of $D$ (data points) 
are obtained by fitting the escape time distribution obtained from
simulations to the distribution function given by Eq. (\ref{eq:pt}) using the
$\beta k$ values as given in Fig. \ref{fig4}. In (a), the dependence of $D$ on
$L$ is fitted by the function of $D=D_\infty+a_\mathrm{L} L^{-2}$ (solid and
dashed) with $D_\infty$ and $a_\mathrm{L}$ the fitting parameters, for $L\leq
110$~{\AA} and for the two temperatures as indicated.
In (b), the fitting curves are obtained by using Eq. (\ref{eq:mut}) and the
corresponding fitting functions found in (a).
} 
\label{fig5}
\end{figure}

By fitting the distribution of the escape time obtained from the simulations to
that given by Eq. (\ref{eq:pt}) with the values of $\beta k$ as given in Fig.
\ref{fig4}, one obtains the effective diffusion constant $D$ of the protein at
the tunnel for $L \leq 110$~{\AA}. Fig. \ref{fig5}a shows that $D$ decreases
with $L$. This dependence reflects the facts that the protein has a changing
shape when escaping from the tunnel, and that the shape depends on $L$.
When $L$ is increased, the initial conformation of the full-length protein at
the tunnel becomes more extended leading to a slower diffusion.
Fig. \ref{fig5}b shows that the mean escape time increases with $L$. As
indicated by Eq. (\ref{eq:mut}), the growth of the escape time on increasing
$L$ is due to both the longer diffusion distance (which is equal to $L$) and
the slower diffusion speed. Fig. \ref{fig5}b also shows that for $L >
110$~{\AA}, the escape time increases with $L$ much faster than for $L \leq
110$~{\AA}, in consistency with the change in diffusion properties shown in
Fig.  \ref{fig4}.

\begin{figure}
\includegraphics[width=3.4in]{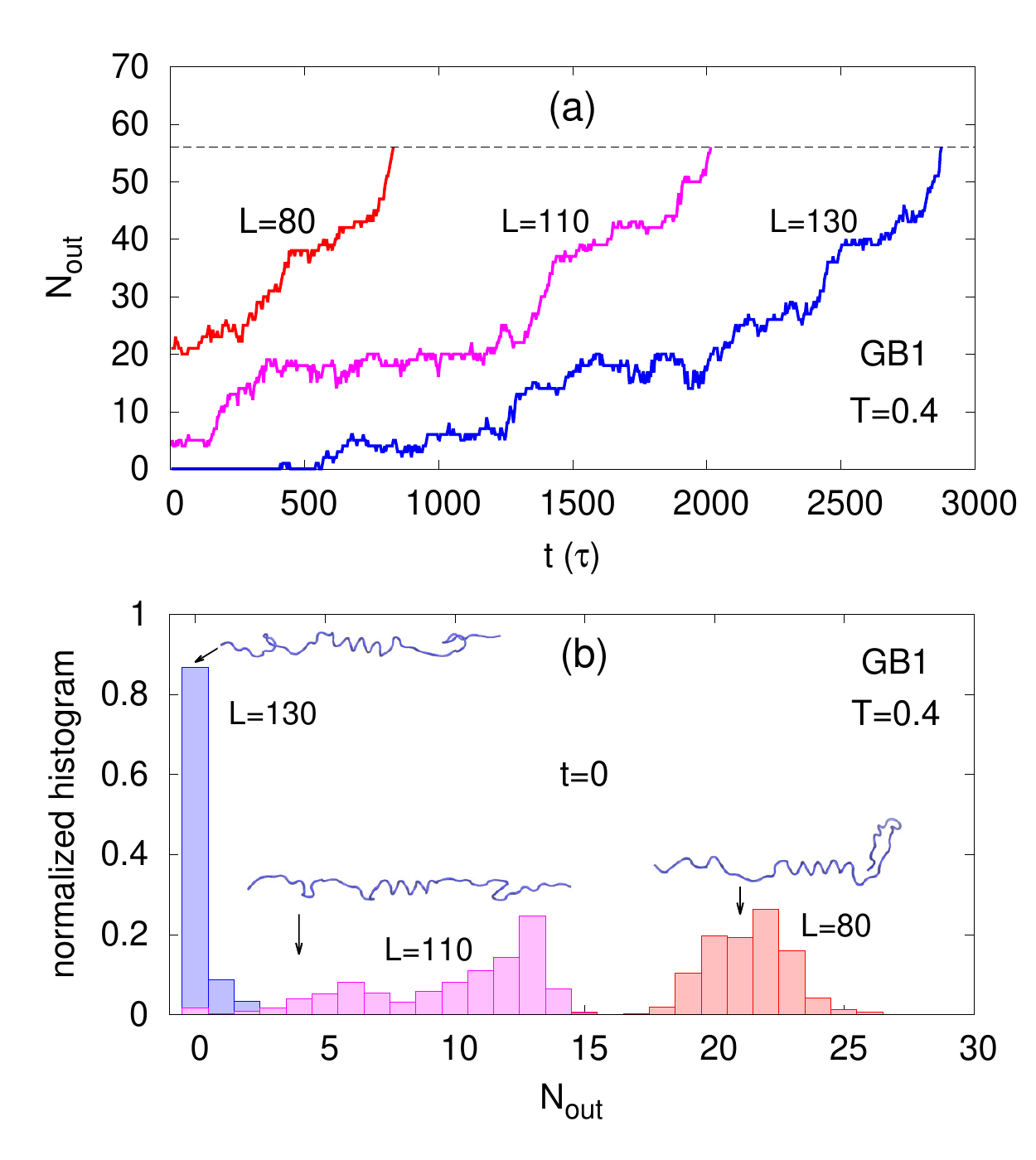}
\caption{
	\TXH{(a) Dependence of the number of amino acid residues outside the
	tunnel, $N_\mathrm{out}$, on the time, $t$, in typical escape
	process of protein GB1 at $T=0.4\,\epsilon/k_B$ for the
	tunnel length of $L=80$ {\AA} (red), $L=110$ {\AA} (magenta) and
	$L=130$ {\AA} (blue), as indicated. The processes are complete when
	$N_\mathrm{out}=56$ (dashed). 
	(b) Histograms of $N_\mathrm{out}$ at the moment the full-length protein
	begins the escape process ($t=0$), obtained from multiple simulations
	of the growth process for the three tunnel lengths as
	considered in (a). The protein conformations at $t=0$ corresponding to
	the trajectories shown in (a) are shown as insets.
	}
	} 
\label{figcrossover}
\end{figure}

The cross-over in the diffusion properties and the escape time observed
at $L\approx 110$~{\AA} for GB1 is related to the relative size of a tunnel
compared to that of a protein. If the tunnel length is such that the protein,
presumably with most of the secondary structures formed, can be found
completely inside the tunnel, then the escape of the protein is much slower
than the case of a shorter tunnel length, in which the protein cannot fit
itself entirely in the tunnel. The tunnel length of 110~{\AA} thus is related
to the size of GB1, such that it can merely have a small part outside the
tunnel at the moment the chain is released from the PTC. The escape process is
accelerated only by the folding of the escaped part of the protein at the
tunnel. \TXH{Fig. \ref{figcrossover}a shows that in typical escape processes,
the number of amino acid residues escaped from the tunnel, $N_\mathrm{out}$,
has similar trends in the time evolution for different tunnel lengths $L$,
except that $N_\mathrm{out}$ has different values at $t=0$, the moment a
full-length protein begins the escape process.
Fig. \ref{figcrossover}b shows that the distribution of $N_\mathrm{out}$
at $t=0$ strongly depends on $L$.  For $L=130$ {\AA}, the protein is mostly
found completely inside the tunnel, i.e. $N_\mathrm{out}=0$. 
On the other hand, for $L=80$ {\AA}, the protein
always has a significant part outside the tunnel with $N_\mathrm{out}$ 
essentially ranging from 18 to 26. For the cross-over length $L=110$ {\AA},
$N_\mathrm{out}$ varies between 0 and 15. The cross-over length approximately
corresponds to the smallest tunnel length for which $N_\mathrm{out}$ can have a
zero value.
}

In consistency with the above mechanism, we find that a similar cross-over of
the diffusion properties and the escape time on increasing the tunnel length is
observed for the three-helix bundle protein SpA.
For SpA, the cross-over occurs at $L \approx 90$~{\AA} (see
Figs. S4 and S5 of supplementary material), quite close to the cross-over
length for GB1, and is consistent with the fact the both proteins are single
domain and of similar size.
The shorter cross-over length for SpA is a little shorter than for GB1
due to the fact that SpA can form more $\alpha$-helices inside the tunnel,
leading to a shorter size than GB1.  Interestingly, the real length of
ribosomal exit tunnel falls between 80~{\AA} and 100~{\AA}, very close to our
estimates of the cross-over length \TXH{for the GB1 and SpA proteins.
Note that the latter are among the smallest single domain proteins.}
It is suggested that the ribosome's tunnel length has been selected
to facilitate an efficient escape of \TXH{small} single domain proteins. 
\TXH{Our study indicates that the cross-over tunnel length increases with the
protein size, thus large proteins would have no problem of escaping the
ribosomal tunnel from the viewpoint of diffusibility.}

\subsection{Effect of macromolecular crowding}

\begin{figure}
\includegraphics[width=3.4in]{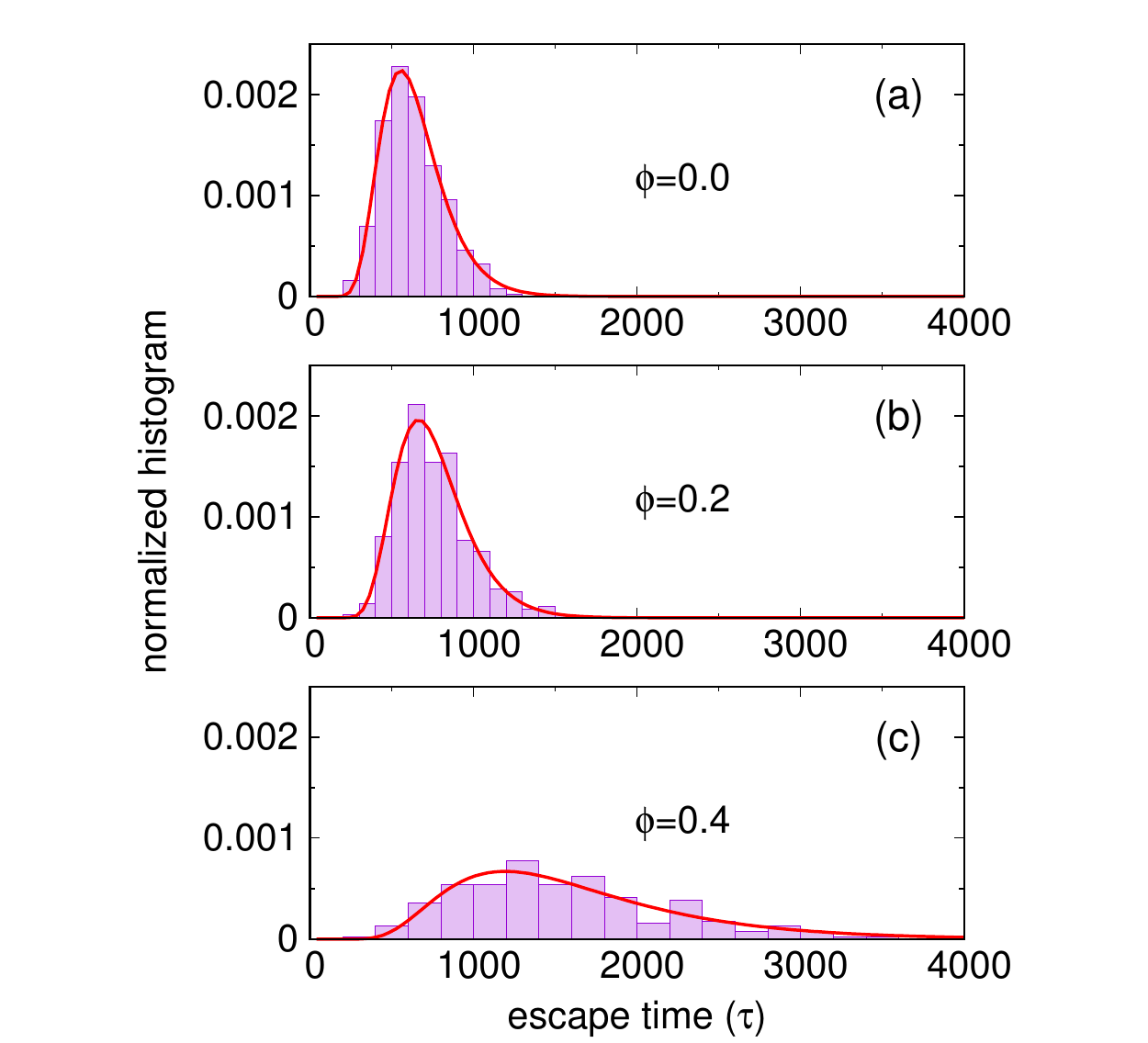}
\caption{Distribution of protein escape time without (a) and in presence of
crowders at the volume fraction $\phi=0.2$ (b) and $\phi=0.4$ (c). The
histograms are obtained for protein GB1 with the C3 contact map with repulsive
crowders at $T=0.8\,\epsilon/k_B$ for the tunnel length of $L=80$~{\AA}.} 
\label{fighisto}
\end{figure}

\begin{figure}
\includegraphics[width=3.4in]{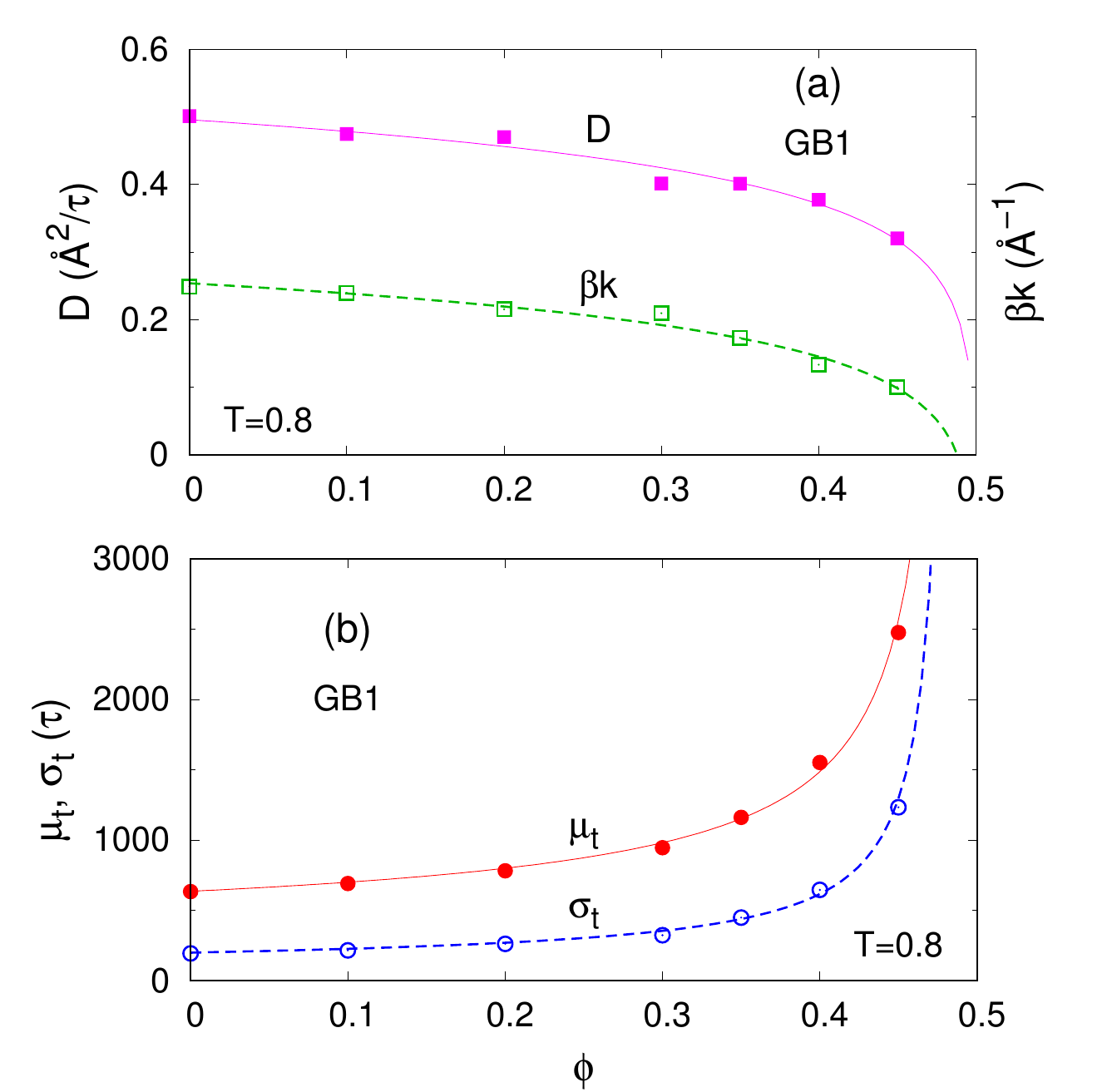}
\caption{(a) Dependence of the diffusion constant, $D$, (filled squares) and
the potential parameter, $\beta k$, (open squares) on the volume fraction
$\phi$ of crowders. (b) Dependence of the mean, $\mu_t$, (filled circles) and
the standard deviation (open circles) $\sigma_t$, of the escape time on $\phi$. 
The data shown are obtained for the GB1 protein with C3 contact map at
$T=0.8\,\epsilon/k_B$ for the tunnel length $L=80$~{\AA} with repulsive
crowders. The values of $D$ and $\beta k$ are obtained by fitting the escape
time distribution from the simulations to that of the diffusion model. The fits
in (a) (solid and dashed lines) have a logarithmic dependence on $\phi$ (see
text), whereas the smooth lines in (b) are calculated from the fitting
functions shown in (a).
}
\label{fig7}
\end{figure}

We proceed now study the escape of nascent protein in the presence of a crowd
of macromolecules outside the ribosomal tunnel. For this investigation, we fix
the tunnel length to be $L=80$~{\AA} and consider various volume fractions
$\phi$ of the crowders. A snapshot of an escaping protein molecule entering
the solution of crowders is shown in Fig. S6 of supplementary material.

First, we consider the case in which the interaction between the crowders and
amino acids are purely repulsive. Fig. \ref{fighisto} shows histograms of the
escape time obtained by the simulations for the GB1 protein at
$T=0.8\,\epsilon/k_B$ for the crowders' volume fraction $\phi=0$, 0.2 and 0.4.
It is shown that as $\phi$ increases the histogram is more spread and shifted
toward higher time values, meaning that the escape time is longer and more
disperse in the presence of crowders. The histograms of the escape time
are found to be consistent with the distribution function given by Eq.
(\ref{eq:pt}) of the diffusion model. The fits to this function give us the
effective values of $D$ and $\beta k$ for different crowder concentrations. 

Fig. \ref{fig7}a shows that both $D$ and $\beta k$ decrease with $\phi$. We
find that both $D$ and $\beta k$ can be approximately described with a
logarithmic dependence on $\phi$ in the following forms: 
\begin{eqnarray}
D & = & D_0 + a \ln \left(1 - \frac{\phi}{\phi_c}\right) \ , 
\label{eq:D} \\
\beta k  & = & \beta k_0  + b \ln \left(1 - \frac{\phi}{\phi_c} 
\right) \ ,
\label{eq:betak}
\end{eqnarray}
where $D_0$ and $k_0$ are the values of $D$ and $k$, respectively, at $\phi=0$;
$a$ and $b$ are the fitting parameters; $\phi<\phi_c$ and $\phi_c$ is a cut-off
volume fraction, beyond which the full escape of the protein becomes
impossible. We find that $\phi_c=0.5$ is a good estimate. The above logarithmic
dependences suggest that the effect of the crowders on the escape process of
protein has an entropic origin, as $(1-\phi/\phi_c)$ can be considered as
the effective volume fraction accessible to the escaping protein in the space
outside the tunnel. Note that entropy loss due to excluded volume is also the
primary effect of crowding and confinement on protein stability
\cite{Minton2001,Zhou2008}. Having the functions given in Eqs. (\ref{eq:D},
\ref{eq:betak}), one can calculate the mean and the standard deviation of the
escape time from Eqs. (\ref{eq:mut},\ref{eq:sigt}) of the diffusion model. Fig.
\ref{fig7}b shows that the mean escape time and the dispersion of the escape
time obtained from simulations at various $\phi$ also agree with the diffusion
model.

\begin{figure}
\includegraphics[width=3.4in]{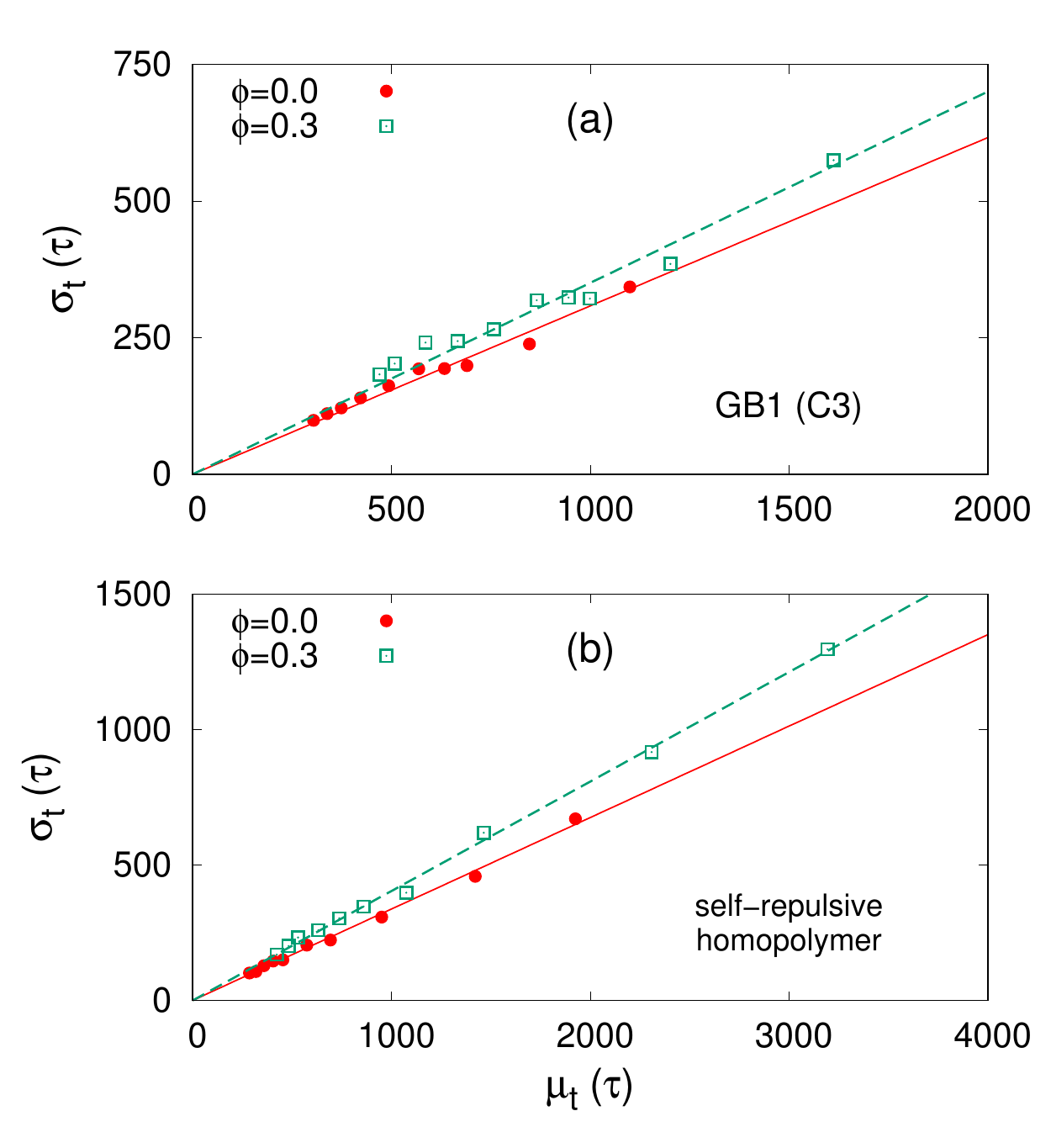}
\caption{Dependence of the standard deviation ($\sigma_t$) on the mean ($\mu_t$)
of the escape time in the cases without crowders (circles) and with repulsive
crowders at volume fraction $\phi=0.3$ (squares) for the GB1 protein with the
C3 native contact map (a) and for the self-repulsive homopolymer (b).
The data points, obtained for various temperatures between 0.3 and 2
$\epsilon/k_B$ for the tunnel length of $L=80$~{\AA}, are fitted by a linear
function for $\phi=0$ (solid) and $\phi=0.3$ (dashed).
	\TXH{
	The fits correspond to $\beta k = 0.264$ {\AA}$^{-1}$ and 0.204
	{\AA}$^{-1}$ for GB1, and $\beta k =0.219$ {\AA}$^{-1}$ and 0.153
	{\AA}$^{-1}$ for the self-repulsive polymer, at $\phi=0$ and
	$\phi=0.3$, respectively.}
	} \label{fig8}
\end{figure}

Fig. \ref{fig8} shows that the standard deviation of the escape time,
$\sigma_t$, depends almost linearly on the the mean escape time, $\mu_t$, for
both the protein and the self-repulsive homopolymer at various temperatures,
indicating that $\beta k$ is constant for each system on changing the
temperature. The value of $\beta k$ however depends on the volume fraction
$\phi$ of the crowders, as indicated by the slopes of the fits shown in Fig.
\ref{fig8}. One finds that $\beta k$ decreases when $\phi$ increases
from 0 to 0.3 for both the protein and the polymer, \TXH{indicating
that diffusion is slower in the presence of crowders. Again here, one 
also finds that for both cases, with and without crowders, the value of $\beta
k$ for GB1 is larger than for the self-repulsive homopolymer, confirming the
enhancing effect of folding on the escape of protein.}

\begin{figure}
\includegraphics[width=3.4in]{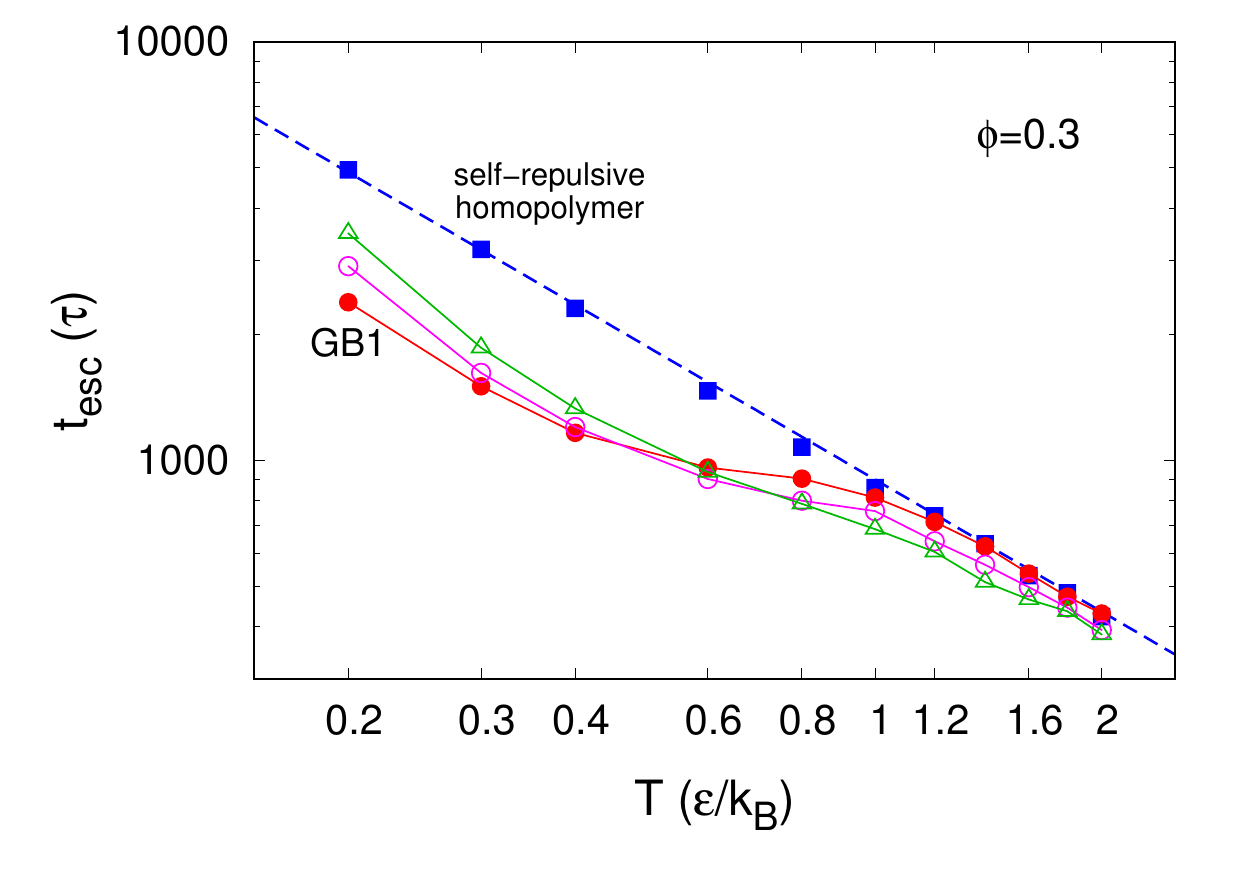}
\caption{
Log-log dependence of the median escape time, $t_\mathrm{esc}$, on temperature,
$T$, in presence of crowders. The data points shown are for a self-repulsive
homopolymer with repulsive crowders (squares), the GB1 protein with repulsive
crowders (filled circles), GB1 with attractive crowders with
interaction strength $\epsilon_1=0.3\epsilon$ (open circles), and GB1 with
attractive crowders with $\epsilon_1=0.5\epsilon$ (triangles). The C3 native
contact map is used for GB1. In all cases, the tunnel length is $L=80$ {\AA}
and the crowders' volume fraction is $\phi=0.3$. The escape times of the
homopolymer are fitted by a straight line with a slope equal to $-1.05$.
}
\label{fig9}
\end{figure}

Fig. \ref{fig9} shows the dependence of the median escape time on temperature
for the protein GB1 and the self-repulsive homopolymer in the presence of
repulsive crowders at the volume fraction $\phi=0.3$. It is shown that
the log-log plot of this dependence for both the protein and the homopolymer
has similar characteristics to that found in Fig. \ref{fig3} for the case
without crowders, except that the escape times are longer with the crowders.
For the homopolymer, the escape time decreases with temperature linearly in the
log-log plot with a slope close to $-1$, indicating that the diffusion
constant of the polymer also depends linearly on temperature like for the case
without crowders. In the presence of repulsive crowders, the escape times of
protein at high temperatures are close to those of the self-repulsive
homopolymer. At low temperatures, favorable for folding, the protein has
significant shorter escape times than the polymer, indicating that the impact
of folding on the escape time is not affected by the crowders. 

\begin{figure}
\includegraphics[width=3.4in]{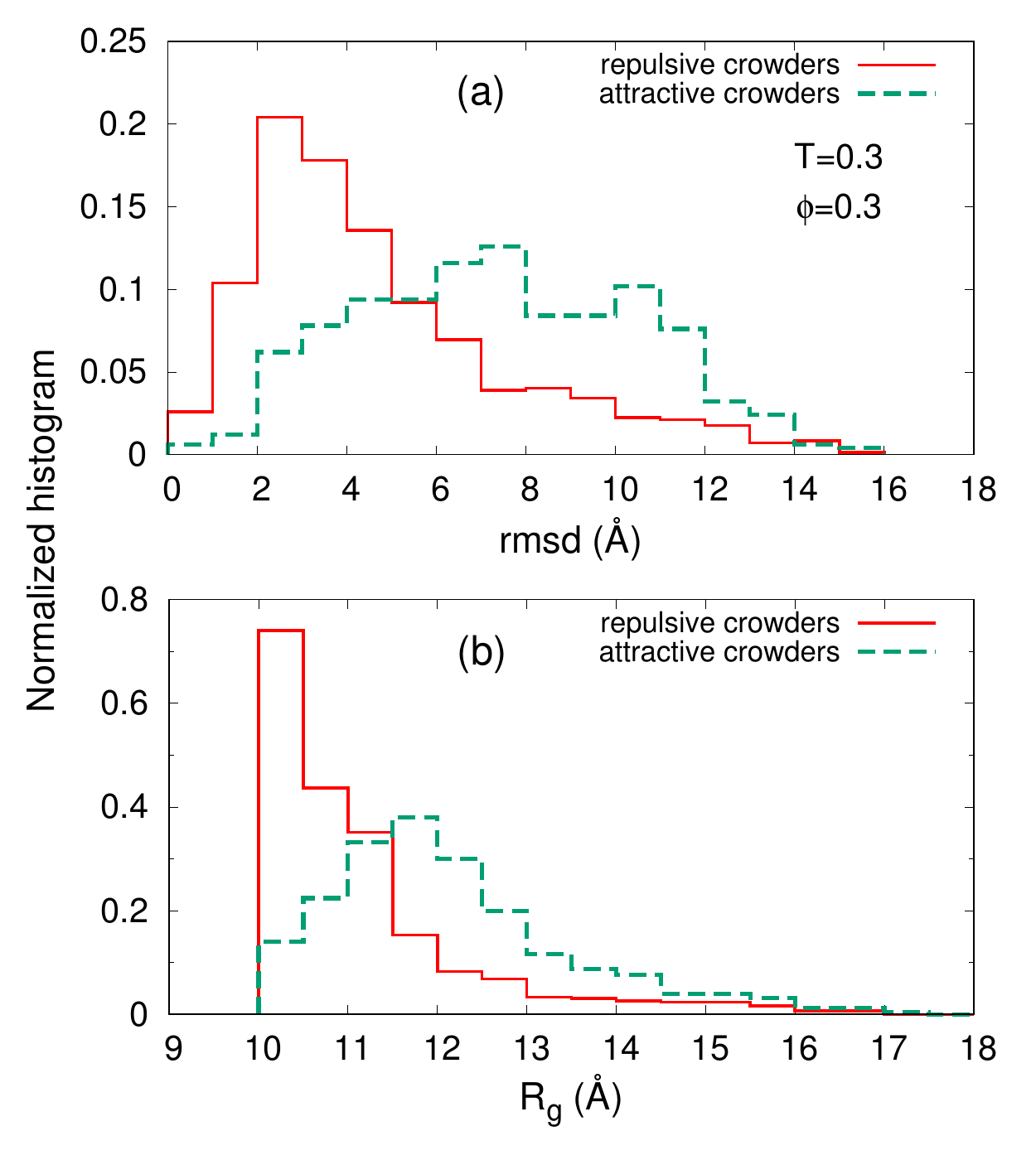}
\caption{
Histogram of the root mean square deviation (rmsd) from the native state (a)
and the radius of gyration ($R_g$) (b) of the protein conformations at the
moment of full escape from the exit tunnel. The data shown are obtained from 
500 independent simulation trajectories at $T=0.3\,\epsilon/k_B$ for protein
GB1 with either repulsive crowders (solid) or attractive crowders (dashed) at
volume fraction $\phi=0.3$. The attractive crowders have the interaction
strength of $\epsilon_1=0.3\epsilon$.
}
\label{fig10}
\end{figure}

Fig. \ref{fig9} also shows the escape times of protein in the presence of
attractive crowders with two different interaction strengths, $\epsilon_1=
0.3\epsilon$ and $\epsilon_1=0.5\epsilon$, of the attraction between crowder
and amino acid. It can be seen that the attractive crowders make the escape
faster than the repulsive crowders, but only at high and intermediate
temperatures. At low temperatures ($T \leq 0.4\,\epsilon/k_B$), the protein
escapes more slowly in the presence of attractive crowders than of the
repulsive ones. Furthermore, the escape time also increases when the
attraction strength $\epsilon_1$ increases at low temperatures. The reason for
this increase is that, in contrast to repulsive crowders, attractive crowders
destabilize the native interactions in protein. Thus, folding is less favorable
in the presence of attractive crowders leading to a weaker enhancement of
folding on the escape speed. Fig. \ref{fig10} shows that the distributions of
the root mean square deviation (rmsd) from the native state and the radius of
gyration of protein conformations obtained at the moment of full escape from
the tunnel are shifted towards higher values when switching from repulsive
crowders to attractive crowders. 

It is believed that ribosome-associated chaperones, such as the trigger factor
(TF) in prokaryotes or the Hsp70 Ssb and NAC (nascent chain-associated complex)
in eukaryotes, are of particular importance in guiding nascent proteins to fold
correctly \cite{Deuerling2005}. The binding of these chaperones to the ribosome
effectively leads to a very high concentration of chaperones near the exit
tunnel \cite{Witt2009}, promoting their interaction with nascent polypeptide.
As a result, the chaperones quickly bind to unfolded, hydrophobic segments of
the polypeptide before these segments can fold or misfold, keeping the nascent
chain unfolded. Our simulations with attractive crowders show a similar effect,
as indicated in Fig. \ref{fig10}, that attractive crowders make the fully
released protein conformation less native-like and less compact than in the
case with repulsive crowders. Our simulations predict that the escape time of
nascent protein increases in the presence of chaperones due to both their
crowding effect and their attractive interaction with hydrophobic segments.

\section{Conclusion}

Post-translational escape of nascent protein at the ribosome is a stochastic
process governed by protein native interactions, the geometry of the
ribosomal exit tunnel and macromolecular crowders outside the tunnel. We
have shown that \TXH{non-local} native interactions
speed up the escape process at temperatures favorable for folding,
\TXH{while the local interactions responsible for the formation of
$\alpha$-helices slow it down. As a consequence, proteins with 
a content of $\beta$-sheets tend to
escape faster than those with only $\alpha$-helices in the native state.}
Increasing the tunnel length or the concentration of crowders also slows down
the protein escape. In the view that the concomitant folding and escape
of nascent protein at the exit tunnel are beneficial for both
the productivity of the ribosome and the protection of nascent protein against
aggregation, it can be conjectured that the protein synthesis machinery has
been evolved to facilitate both the folding and the escape of nascent proteins.
In support of this conjecture, we have shown that real ribosomal exit tunnel
has adopted the length that is close to a cross-over length of the tunnel,
beyond which the protein escape falls into a regime of a much slower diffusion
for \TXH{small} single domain proteins. 

Our study shows that repulsive crowders outside the tunnel induce an entropic
effect on the diffusion properties of protein at the tunnel, leading to
increased escape times but does not change the enhancing effect of folding on
the escape process. The latter effect is changed only in
the case of attractive crowders, whose attraction to amino acids competes with
native interactions in the nascent polypeptide. Due to this competition the
fully escaped protein conformation is more extended and less native-like. The
unfavorable effect of attractive crowders on the folding of nascent protein is
also reflected on the increased escape times at low temperatures, as shown in
our study. It is suggested that the ribosome-associated chaperones induce
similar effects on nascent polypeptides as found with attractive crowders. 

Low-dimensional diffusion models have been successfully applied to study
complex dynamics \cite{Kramers1940,Zwanzig1988,Peters2013}.
Our work proves that the simple diffusion model considered is useful for
understanding the escape of protein at the exit tunnel. The results suggest
that intrinsically disordered proteins, considered as the self-repulsive
homopolymer in our study, have longer escape time than \TXH{foldable proteins
with a significant number of long-range contacts},
and their diffusion is the most akin to that of a Brownian particle.

This research is funded by Vietnam National Foundation for Science and
Technology Development (NAFOSTED) under Grant No. 103.01-2016.61.

\section*{Supplementary Material}
See supplemental material for the specific heats for the Go-like model of GB1
with different native contact maps, the dependence of the escape
time on temperature for the helical protein 2rjy and a single
$\alpha$-helix, the dependence of the dispersion of the escape time on the
mean escape time for SpA, the dependence the diffusion constant and the mean
escape time on the tunnel length for SpA, and a snapshot of an escaping protein
entering a solution of macromolecular crowders.

\bibliography{refs_tunnel}

\clearpage

\setcounter{equation}{0}
\renewcommand\theequation{S\arabic{equation}}

\setcounter{figure}{0}
\renewcommand\thefigure{S\arabic{figure}}

\setcounter{table}{0}
\setcounter{page}{1}

\onecolumngrid

\makeatletter

\section*{Supplementary Material for:
``Protein escape at the ribosomal exit tunnel: effects of native interactions,
tunnel length and macromolecular crowding'', P.T. Bui and T.X. Hoang}

\begin{figure}[!ht]
\includegraphics[width=3.4in]{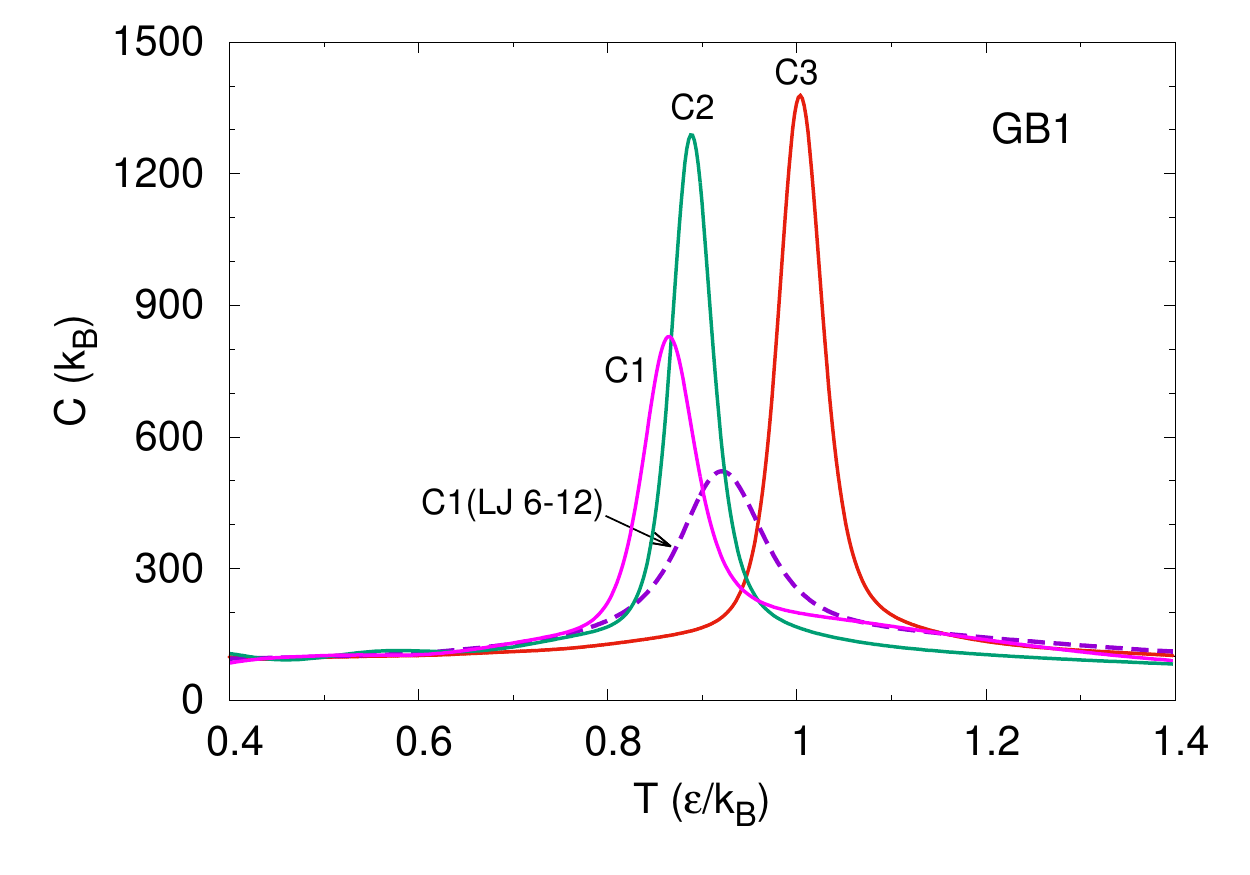}
\caption{
Dependence of the specific heat, $C$, on the temperature, $T$, for the GB1
protein in Go-like models with three different native contact maps, C1
(magenta), C2 (green) and C3 (red), with the 10-12 Lennard-Jones (LJ) potential
for the
native contacts (see Methods section of the main text); and for the C1 contact
map with the 6-12 LJ potential for native contacts (dashed).
The folding temperature $T_f$, defined as the temperature of the
maximum of the specific heat, is equal to 0.866, 0.888 and 1.004 $\epsilon/k_B$
for the models with the C1, C3, and C3 maps, respectively, with the 10-12
potential; and 0.922 $\epsilon/k_B$ for the model with the 6-12 potential.
}
\label{fig:s1}
\end{figure}

\begin{figure}[!ht]
\includegraphics[width=3.4in]{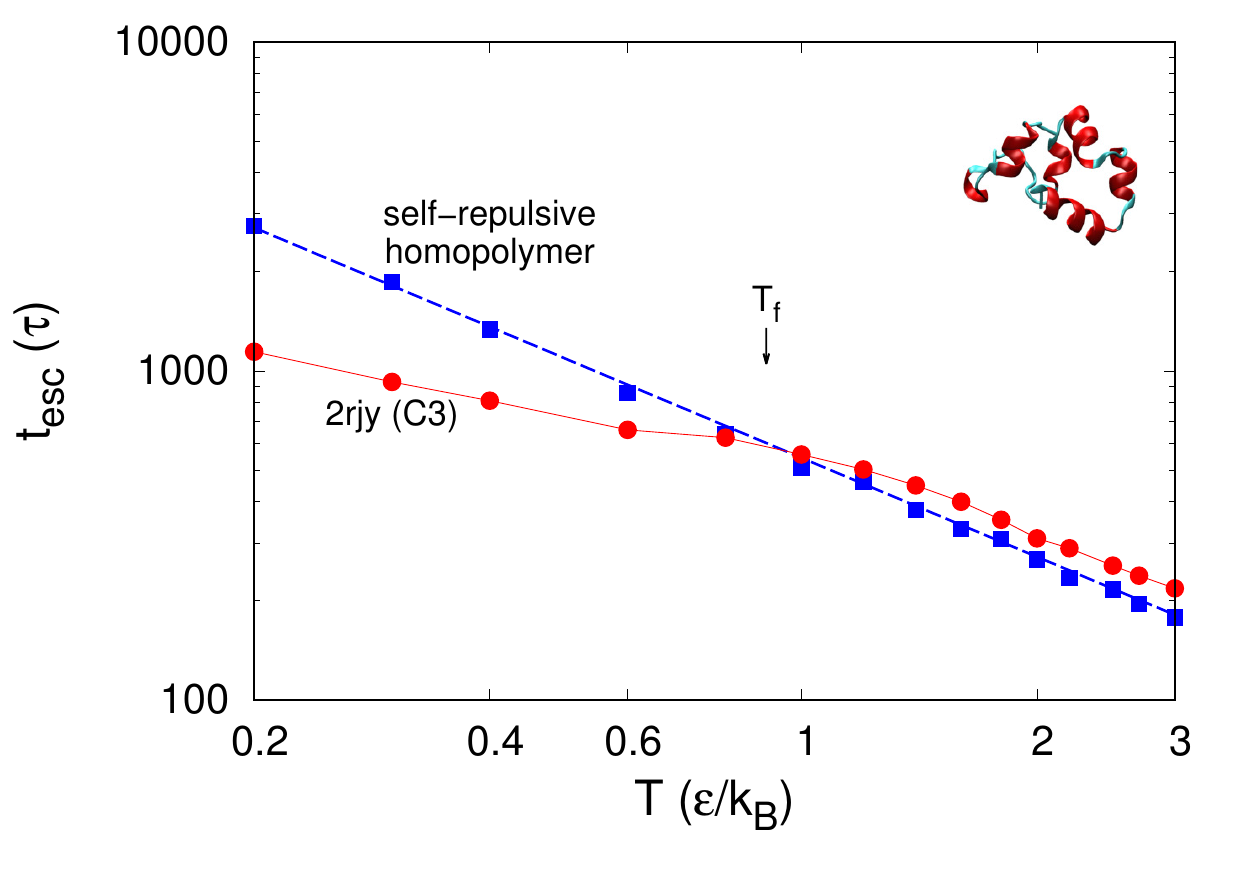}
\caption{
Dependence of the median escape time, $t_\mathrm{esc}$, on the temperature, $T$,
for the villin headpiece protein (PDB code: 2rjy) (circles), a helical
protein of length $N=64$ amino acids, and a same-length self-repulsive
homopolymer (squares). The escape times are calculated for the tunnel
of length $L=80$ {\AA} and without the crowders. The protein is considered in
the Go-like model with the C3 native contact map.  The protein native state is
shown as inset and the arrow indicates its folding temperature $T_f$.
}
\label{fig:s2}
\end{figure}

\begin{figure}[!ht]
\includegraphics[width=3.4in]{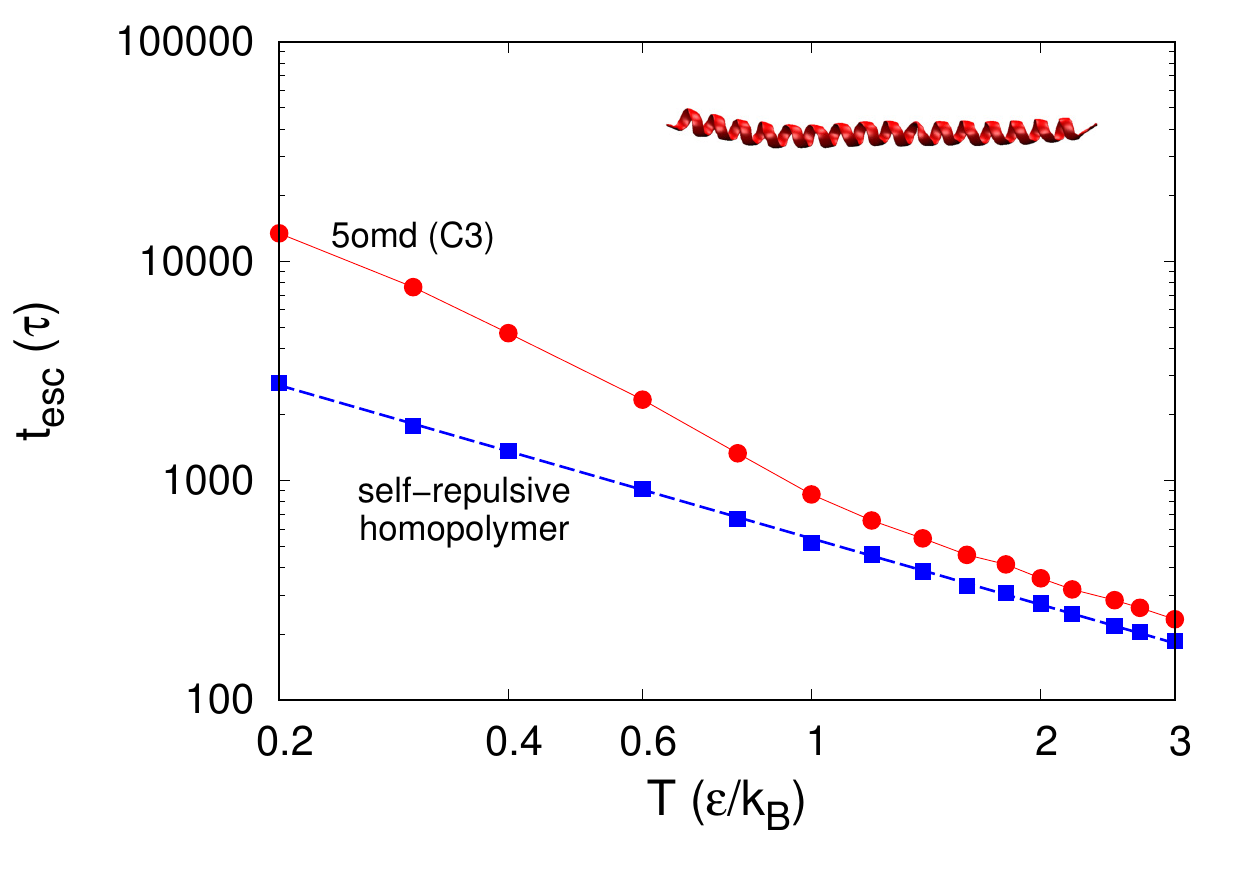}
\caption{
Same as Fig. S2 but for a single $\alpha$-helix of length $N=66$ amino acids
from the S. cerevisiae Ddc2 N-terminal coiled-coil domain (PDB code: 5omd).
}
\label{fig:s3}
\end{figure}

\begin{figure}[!ht]
\includegraphics[width=3.4in]{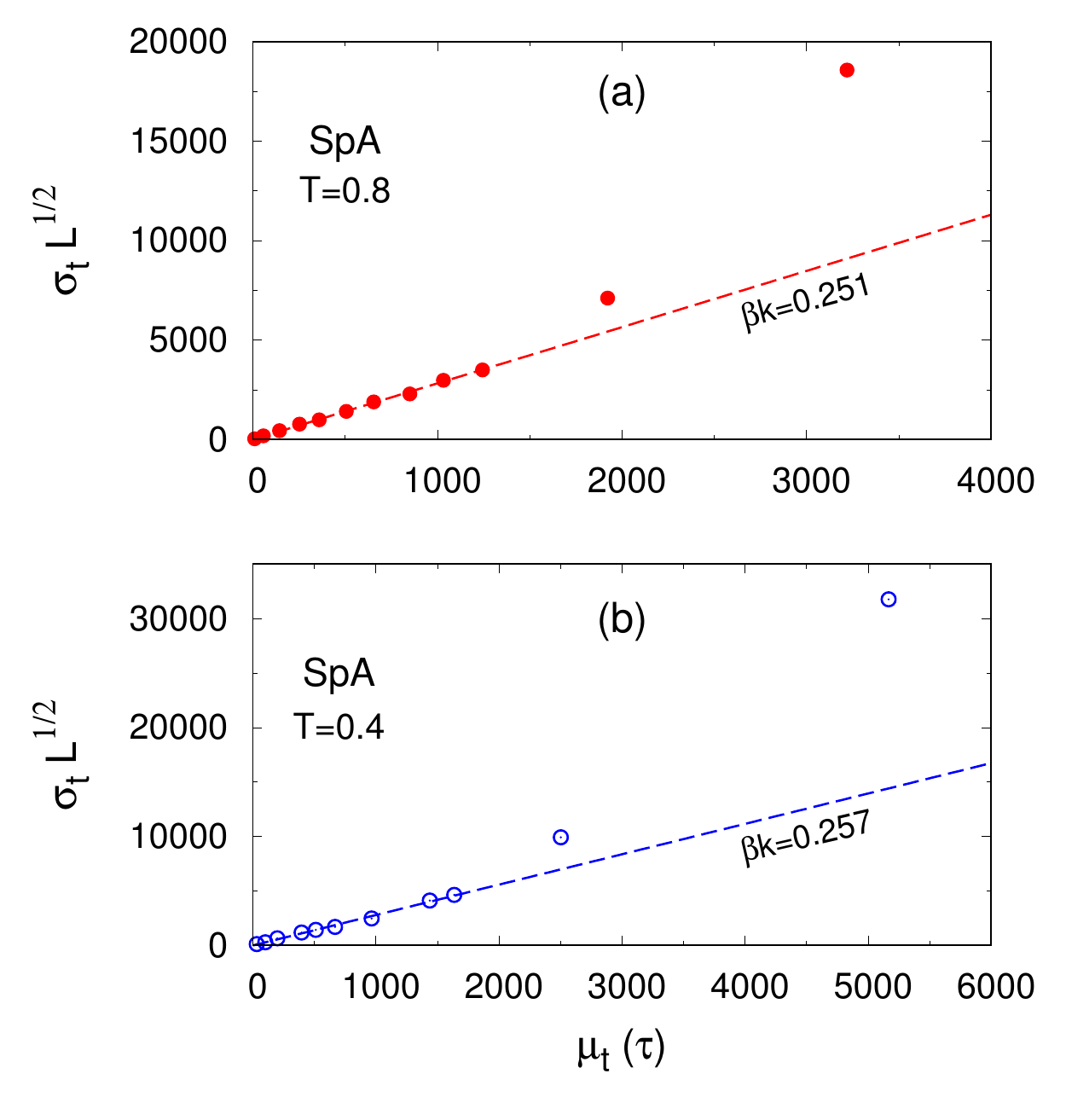}
\caption{
Dependence of the standard deviation of the escape time, multiplied
by the square root of the tunnel length, $\sigma_t L^{1/2}$, on the mean escape
time, $\mu_t$, for the Z domain of Staphylococcal protein (SpA) in the Go-like
model with the C3 native contact map. The panels show the simulation data
for two temperatures, $T=0.8\,\epsilon/k_B$ (a) and $T=0.4\,\epsilon/k_B$ (b).
Different data points corresponding to different tunnel length $L$, 
considered to take values between 30~{\AA} and 120~{\AA}. The data points of
$L\leq 100$~{\AA} are fitted to the diffusion model (dashed line) with a
constant $\beta k$, equal to 0.251~{\AA}$^{-1}$ (a) and 0.257~{\AA}$^{-1}$ (b),
as indicated.  }
\label{fig:s4}
\end{figure}

\begin{figure}[!ht]
\includegraphics[width=3.4in]{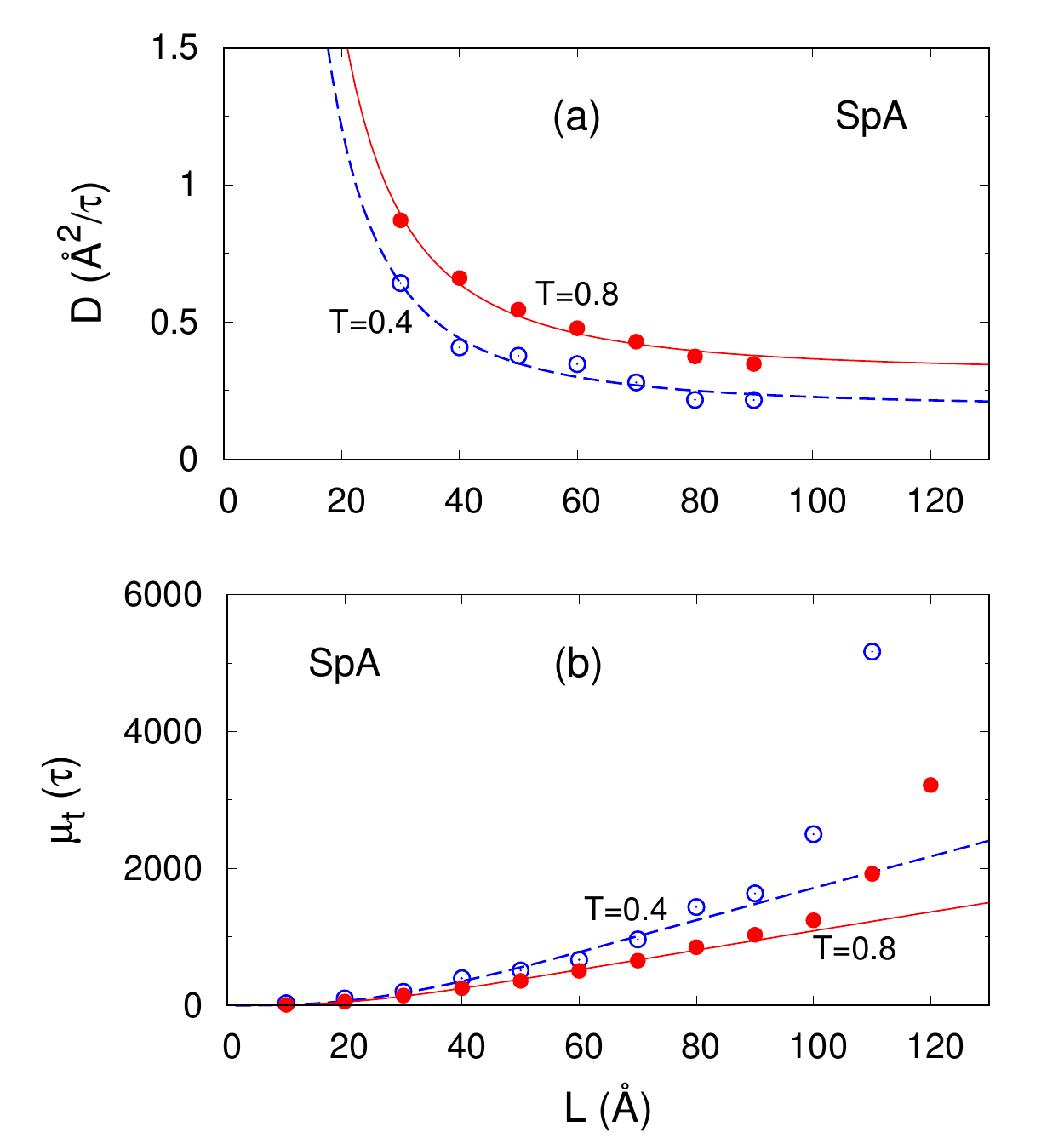}
\caption{
Dependence of the diffusion constant, $D$, (a) and the mean escape time,
$\mu_t$, (b) on the tunnel length $L$, for protein SpA at $T=0.8\epsilon/k_B$
(filled circles) and $T=0.4\,\epsilon/k_B$ (open circles). The values of $D$
for $L$ between 30~{\AA} and 90~{\AA} are obtained by fitting the escape time
distribution obtained from simulations to the diffusion model with the $\beta
k$ values given in Fig. S2. 
The dependence of $D$ on $L$ is fitted by the function $D=D_\infty +
a_\mathrm{L} L^{-2}$, with fitting parameters $D_\infty$ and $a_\mathrm{L}$,
for $T=0.8\,\epsilon/k_B$ (solid) and
$T=0.4\,\epsilon/k_B$ (dashed). The
dependence of $\mu_t$ on $L$ obtained by the diffusion model is shown as solid
and dashed lines for the two temperatures, as indicated.
}
\label{fig:s5}
\end{figure}

\begin{figure}[!ht]
\includegraphics[width=4.5in]{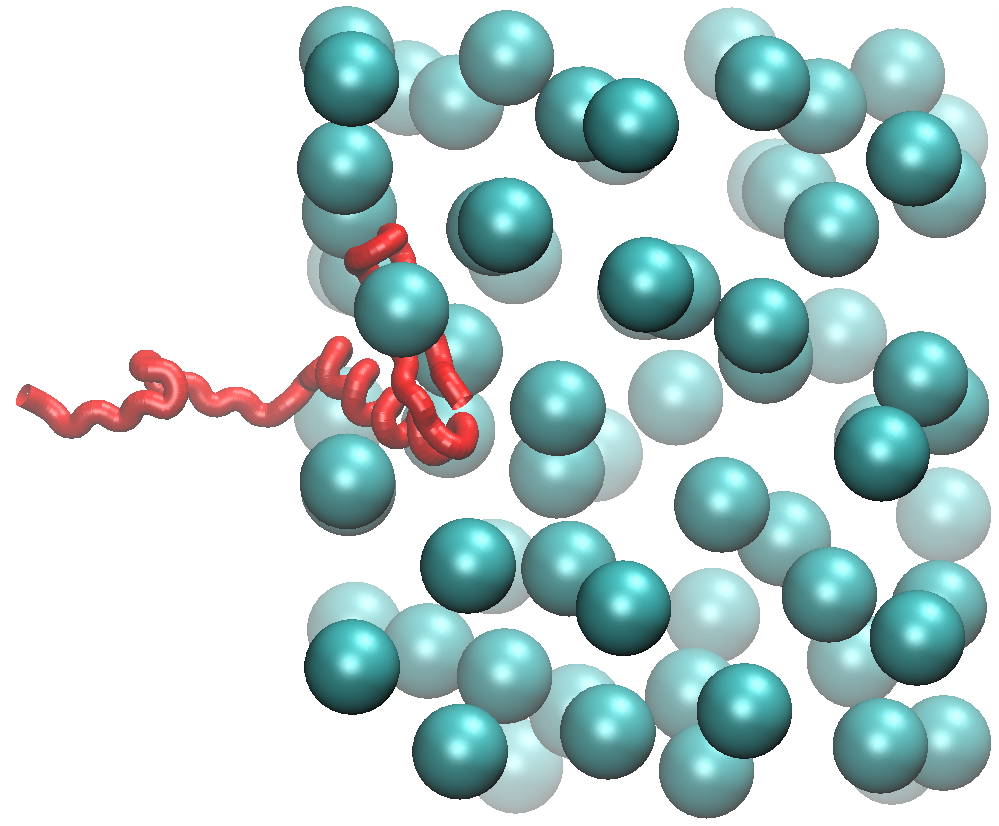}
\caption{
Snapshot of the protein GB1 (red) escaping into a solution of crowders (cyan)
with the crowders' volume fraction $\phi=0.3$.
}
\label{fig:s6}
\end{figure}

\end{document}